\documentclass[10pt,preprint]{aastex}
\usepackage{natbib}
\newcommand{\myemail}{p.cargile@vanderbilt.edu}
\slugcomment{Published in The Astronomical Journal}
\shorttitle{X-ray Properties of Blanco~1}
\shortauthors{Cargile, James, \& Platais}
\begin{document}
\title{A New X-ray Analysis of the Open Cluster Blanco~1 \\ Using Wide-field $BVI_{c}$ Photometric\footnote{Based on observations made with the Small- and Medium-Aperture Research Telescope System ({\sc smarts}) at the Cerro Tololo Inter-American Observatory.}  and Proper Motion Surveys}
\author{P.~A. Cargile\altaffilmark{1},  D.~J. James\altaffilmark{1,2}, and I. Platais\altaffilmark{3}}
\altaffiltext{1}{Department of Physics and Astronomy, Vanderbilt University, Nashville, TN 37235, USA, \myemail}
\altaffiltext{2}{Physics and Astronomy Department, University of Hawai'i at Hilo, Hilo, HI 96720, USA}
\altaffiltext{3}{Department of Physics and Astronomy, Johns Hopkins University, Baltimore, MD, 21218, USA}
\begin{abstract}
We perform a new analysis of the extant {\sc rosat} and {\sc xmm}-Newton X-ray surveys of the southern open cluster Blanco~1, utilizing new $BVI_{c}$ photometric and proper motion data sets. In our study, we match optical counterparts to 47 X-ray sources associated with Blanco~1 cluster members, 6 of which were listed in previous X-ray studies as cluster nonmembers. Our new catalog of optical counterparts to X-ray sources clearly traces out the Blanco~1 main sequence in a color-magnitude diagram, extending from early G to mid-M spectral types. Additionally, we derive new X-ray luminosities as well as ratios of X-ray to bolometric luminosities for confirmed cluster members. We compare these X-ray properties to other young open clusters, including the coeval Pleiades cluster, to investigate the relationship between age and X-ray activity. We find that stars in Blanco~1 generally exhibit X-ray properties similar to those of other open clusters, namely increasing $L_{x}/L_{bol}$ with reducing mass for earlier-type stars, and a saturation limit of $L_{x}/L_{bol}$ at a magnitude of $10^{-3}$ for stars with $V-I_{c} \ga 1.25$. More generally, the X-ray detected stars in Blanco~1 have X-ray emission magnitudes that agree with the overall trends seen in the other young clusters. We observe that X-ray emission decays as a function of age and the rate of this decay is mass dependent. Specifically, for higher mass stars, the trend is Skumanich like (i.e. $L_{x}/L_{bol} \varpropto$ age$^{-1/2}$); however, as one goes to lower masses the magnitude of X-ray emission becomes less of a function of age. In fact, for the lowest mass stars (M-type), there is no observable reduction in X-ray production during the first $\sim$1 Gyr of their lives. However, due to a lack of sensitivity to low X-ray fluxes, there may exist M-type stars that have less than saturated levels of X-ray flux which are not included in our study. In a direct comparison of Blanco~1 to the Pleiades open cluster, members of both clusters have similar X-ray characteristics; however, there does appear to be some discrepancies in the distribution of $L_{x}/L_{bol}$ as a function of color that may be related to scatter seen in the Pleiades {\sc cmd}. Moreover, previous comparisons of this nature for Blanco~1 were not possible due to the reliance on photographic photometry. This is where the power of precise, homogeneous, and standardized {\sc ccd} photometry allows for a high fidelity, detailed study of the X-ray properties of stars in Blanco~1, as well as a thorough comparison of Blanco~1 to other well-studied open clusters.
\end{abstract}
\keywords{open clusters and associations: general --- open clusters and associations: individual (Blanco~1) --- stars: activity --- stars: evolution --- X-rays: stars}
\section{Introduction}\label{Intro}
X-ray emission is a common characteristic among young, main-sequence, solar-type stars. Magnetic fields, produced through a solar-type, magneto-hydrodynamical dynamo process \citep{Parker1955a,Parker1979}, are  able to confine and heat plasma to extreme temperatures  ($\ge 10^{6}$ K) in stellar coronae. This highly energized plasma is responsible for the production of the observed X-ray emission from solar-type stars \citep{Rosner1978, Sams1992}.

Because this plasma heating is highly dependent on the magnetic fields produced by a rotation-induced dynamo, there  is a strong, causal relationship between stellar rotation and magnetic field production, and hence induced magnetic activity (e.g., X-rays). In fact, the efficiency of the stellar dynamo is related to both the rotation period, $P$, and the convective turnover time in the convection zone, $\tau_{c}$, through the Rossby number ($R_{o} = P/\tau_{c}$). Thus for stars with identical rotation periods, we expect increasing magnetic flux levels for stars having increasing $\tau_{c}$ (i.e. decreasing stellar mass) due to a more efficient dynamo. Of course, for decreasing stellar mass, the amount of magnetic flux threading the stellar surface is itself a decreasing function of stellar radius.

Observationally, this framework has proved to be correct,  with a positive correlation existing between coronal emission and rotation rate (or Rossby number). For the most part,  faster rotating, or lower Rossby number, solar-type stars  have higher X-ray luminosities compared with stars with slower rotation rates, or higher Rossby number \citep[e.g.][]{Pallavicini1981,  Vilhu1987, Hempelmann1995}. However,  observations of more rapidly rotating stars (or those with  lower Rossby numbers) are suggestive of a scenario where the  dynamo process, or the magnetic heating rate, does not continue  increasing without limit and appears to saturate.  The data unequivocally show that for late-type stars rotating above $\simeq 15-20$ km~s$^{-1}$, or for M-dwarfs more like $\simeq$ 5-8  km~s$^{-1}$, a saturation plateau of maximal coronal X-ray luminosity occurs at the level of 0.1 \% of the stellar bolometric luminosity \citep[$L_{X}/L_{bol} \sim 10^{-3}$;][]{Vilhu1987, Stauffer1994, James2000}.  These X-ray characteristics are readily seen in observations of relatively young (<1 Gyr) Galactic open clusters.

Open clusters, being natural samples of stars with the same age,  distance, composition, and environmental formation conditions,  have long been considered powerful laboratories to test the models  of stellar formation and evolution in our Galaxy. Through the study  of large numbers of open clusters at several different ages,  the complex, interdependent roles of rotation, age, composition, mass, and initial  conditions play in determining the levels of X-ray emission from  solar-type stars can be investigated.  

Blanco~1 is a relatively nearby, young open cluster (250 pc, $60-100$ Myr; {\sc webda} open clusters database\footnote{The {\sc webda} database, developed by J.-C. Mermilliod, can be found at  http://www.univie.ac.at/webda/}) that is of considerable scientific  interest due to its high Galactic latitude ($b=-79\arcdeg$) and comparable  age to the well studied Pleiades open cluster ($\sim80-120$ Myr; \citealt{Meynet1993}, $[Fe/H] = -0.034 \pm 0.024$;\citealt{Boesgaard1990}). Considerable interest in the cluster has been driven by its reported  metal-rich nature ($[Fe/H] = +0.23$; \citealt{Edvardsson1995}), although  a more recent, self consistent determination now makes the cluster of  near-solar composition ($[Fe/H]=+0.04 \pm 0.04$; \citealt{Ford2005}). The combination of the cluster's systemic velocity (RV$_{sys}$=+5.5 km~s$^{-1}$; \citealt{Mermilliod2008}), its Galactic latitude, and distance below the Galactic plane ($\sim$250~pc) suggests that, if Blanco~1 has an age of $>$50~Myr, it should have been created in or near to the Galactic plane. The unusual location of Blanco~1 makes it unique among the well studied, young ($<<$1~Gyr) open clusters.  

Blanco~1 contains more than 200 known members spread over $\simeq 3\times3$ deg$^{2}$ on the sky \citep{Jeffries1999a,James2009,Mermilliod2008,Moraux2007}. Such a wide areal on-sky distribution  is challenging for deep and complete photometric surveys. The existing X-ray studies of Blanco~1 \citep{Micela1999a, Pillitteri2004} have relied  on older, photographic photometry \citep{deEpstein1985} to segregate cluster members from Galactic  field stars and background galaxies. Due to the  inherent scatter in photographic photometry, especially  in the photometric precision of fainter objects (see \S \ref{OP}), definitive membership status and characteristics of optical counterparts to X-ray sources remain poorly established. 

In this paper, we re-analyze the X-ray properties of Blanco~1 using a recent, standardized {\sc ccd} $BVI_{c}$ photometric survey of the central 1.6$\times$1.3 deg$^{2}$ of Blanco~1. This high fidelity photometric data set, in concert  with a new proper motion survey, allows us to determine a well constrained membership catalog with standardized photometry down to $V\sim17$ (\S \ref{exob}). We combine this new catalog with the available {\sc rosat} and {\sc xmm}-Newton X-ray data to compute accurate X-ray luminosities for those photometric sources identified as cluster members. We utilize our new analysis to examine the X-ray properties of the cluster, including the X-ray luminosity and the ratio of X-ray to bolometric luminosity versus intrinsic color distributions (\S \ref{xrayfl}). Furthermore, we compare these X-ray properties to other well studied open clusters at various ages to investigate how these distributions evolve with time (\S \ref{discusion}).

\section{Extant Observations}\label{exob} 
\subsection{Optical Photometry}\label{OP} 
In order to investigate the photometric membership of Blanco~1, \citet{deEpstein1985} performed a large-scale survey of the central 1.5 deg$^{2}$ of the cluster, utilizing archival photographic plates, obtained at the El Leoncito Observatory in San Juan, Argentina. They were able to identify some 1500 stellar objects down to a limiting magnitude of V$\simeq 16.5$, which corresponds to a late-K spectral type for the reddening and distance of Blanco~1. These authors claim cluster membership for $\simeq 10\%$  of their sample although, as they themselves note, classification of stars fainter than V$=12.6$ must be considered ``tentative''. This is because they produced a photometric dataset based on a calibration between their photographic magnitude system and the then-existing photoelectric dataset for Blanco~1 stars, which unfortunately must be extrapolated for stars fainter than V$\simeq$12.6. 

In light of the paucity of precise and accurate photometric data for Blanco~1, especially for fainter cluster members, a recent study produced a standardized $UBVI_{c}$ {\sc ccd} photometric dataset for the central 1.6$\times$1.3 deg$^2$ of Blanco~1, centered on {\sc ra} (2000): $00^{h} 05^{m}$; {\sc dec} (2000): $-30\arcdeg 02\farcm4$. The details of this survey are given in \citet[][ hereafter J09]{James2009}; here, we merely outline the most pertinent features of their study. $UBVI_{c}$ {\sc ccd} photometric data were taken using the {\sc smarts} 1-m telescope at the Cerro Tololo Inter-American Observatory ({\sc ctio}), equipped with the $19\farcm3$$\times$$19\farcm3$ Y4K camera. The photometric catalog contains 1668 stellar objects with a limiting magnitude of V$\simeq$17. Standardization of their instrumental photometric magnitudes comprised some 60 standard stars nightly, with external errors in transforming instrumental magnitudes onto the standard system of $<2$\%. Comparison of control field photometric data shows that internal errors of their photometric catalog are better than $0.5\%$ (V$<$16) and statistical errors are $\leq$1\% at V$<$16.

\begin{figure}[ht] \epsscale{0.75} \plotone{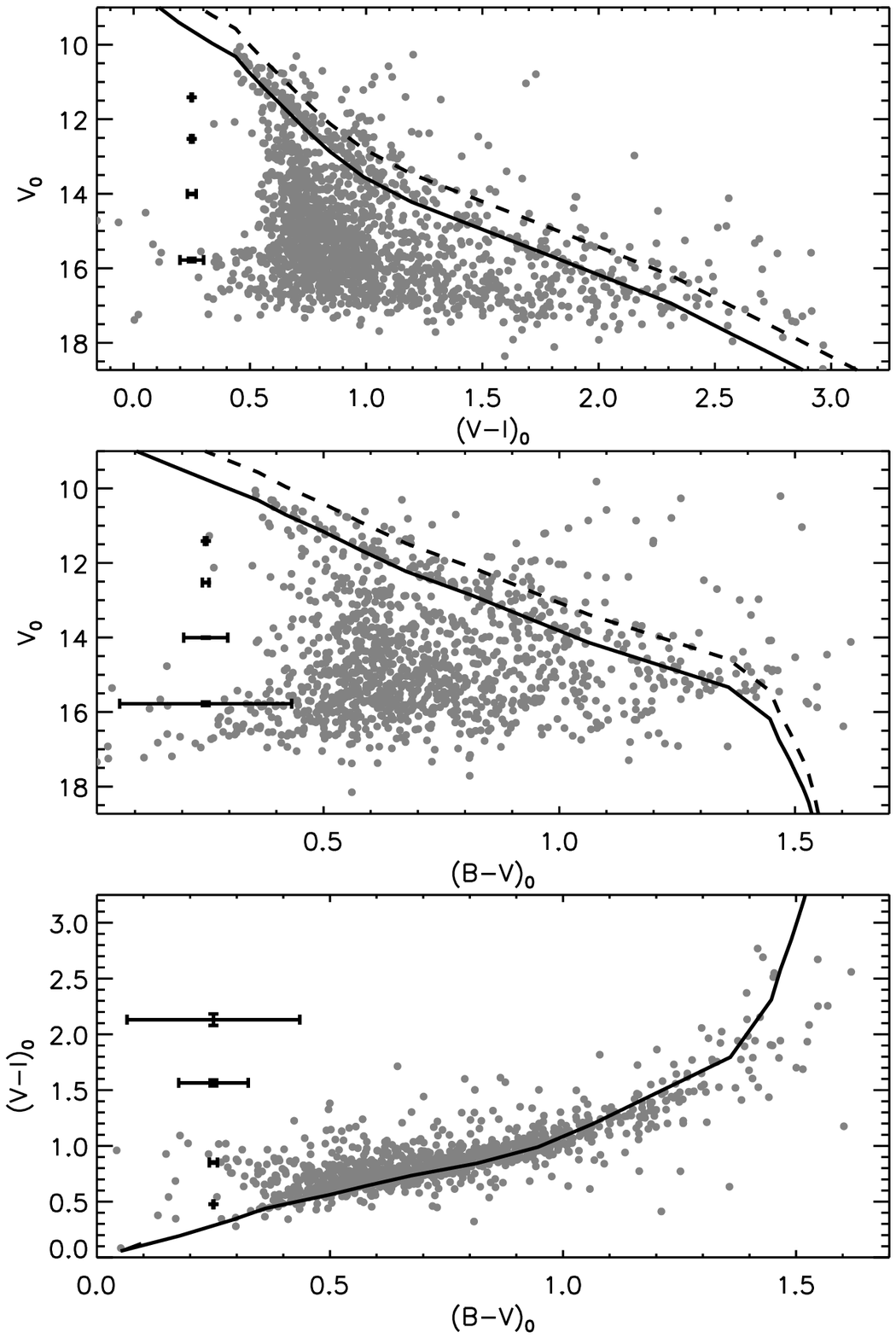} 
   \caption{
     \label{fig1}
     Intrinsic color-magnitude and color-color diagrams for the full J09 photometric catalog are plotted. An $E(B-V) = 0.016$ and $E(V-I_{c})=0.02$ has been used. 
     We overplot the best-fit \citet{DAntona1997} isochrone from the \citet{James2009} paper ({\it solid line}), as well as the equal mass 
     binary sequence ({\it dashed line}). On the left, we plot the average error bars for objects that lie along the main-sequence isochrone. 
   } 
 \end{figure}

Figure \ref{fig1} shows the $V_0$ versus $(B-V)_0$ and $(V-I_{c})_0$ color-magnitude diagrams ({\sc cmd}) based on this photometry. In our study, to convert to intrinsic magnitudes and colors for the $BVI$ {\sc ccd} photometry, we assume a distance of 240 parsecs as derived from isochrone modeling in J09, as well as $E(B-V) = 0.016$ and $E(V-I) = 0.02$ taken from an average of published reddening coefficients for Blanco~1 \citep{Epstein1968,Eggen1970,Eggen1972,Appenzeller1975,Perry1978,deEpstein1985,Westerlund1988}. We note that because Blanco~1 lies significantly out of the Galactic plane, we do not expect differential reddening to be an issue. J09 use a $\tau^{2}$ isochrone fitting routine, as described in \citet{Naylor2006}, to model the main sequence of Blanco~1 with a theoretical isochrone from \citet{DAntona1997}. They find a model-dependent distance and age for the cluster of $240\pm10$ parsecs and $80\pm20$ Myr, respectively. This distance estimate agrees with the distance (242~pc) from the revised Hipparcos parallax for Blanco~1 at $\pi$=4.14$\pm$0.17 mas \citep{vanLeeuwen2007}. We note that fainter than V$\sim$15~mag there is an increased possibility for objects not associated with Blanco~1 to contaminate a photometric membership catalog due to large uncertainties in $B-V$ and $V-I$ (see error bars in Fig. \ref{fig1}). Therefore, it becomes essential to use other membership properties, e.g. proper motions (see \S \ref{pm}), to determine a high fidelity membership list for the faintest stars in Blanco~1.

\subsection{Systematics in Previous Photometry}\label{photscat}
\begin{figure}[ht] \epsscale{1.0} \plotone{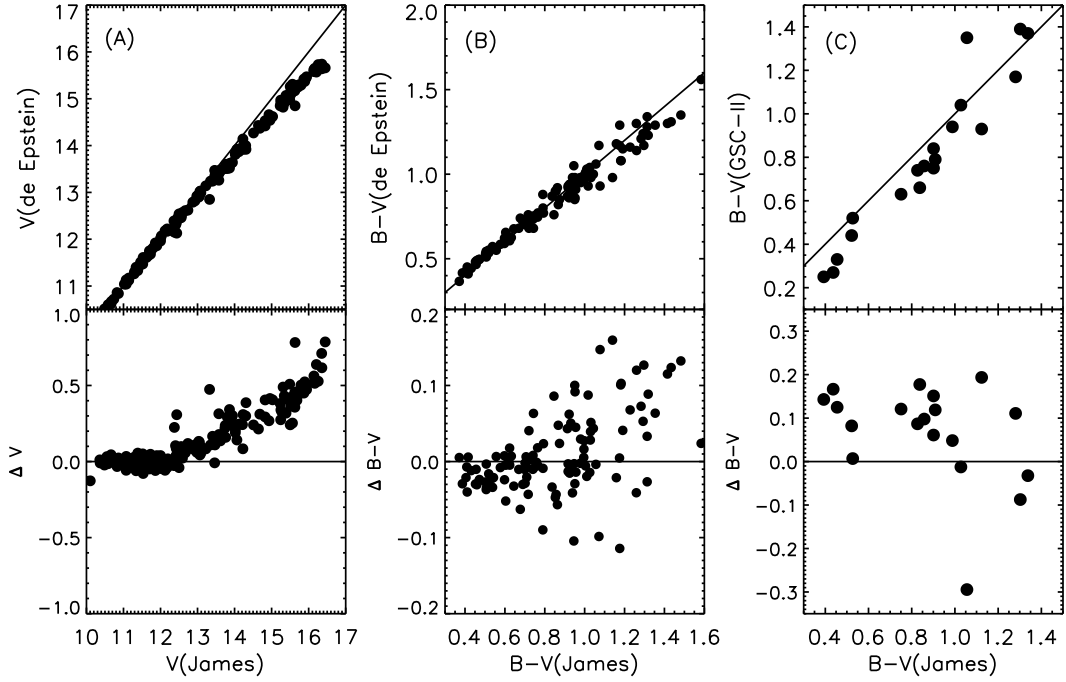}
   \caption{
     \label{fig2}
     $V$ and $B-V$ comparison plots for the \citet{deEpstein1985} (A,B) and {\sc gsc-ii} (C) photometric catalogs with the photometry given in 
     J09 are presented. The solid lines indicate equality between the systems, and are not fits to the data.
   }
 \end{figure}

It is instructive to examine the sets of photometric data used in the analysis of X-ray properties of Blanco~1 members. In Fig.\ \ref{fig2}a,b, a comparison between the photometry of \citet{deEpstein1985} and J09 is plotted. Clearly, for magnitudes greater than V$\sim 12.5$, there is an increasingly large offset between the two catalogs, as well as an increase in the $B-V$ scatter for stars redder than $\sim 0.7$. This scatter is at a $1\sigma$ level of 0.05 for $B-V > 0.7$. These effects are most likely due to the limitations in photographic-to-photoelectric calibration of the de Epstein \& Epstein catalog. 

Employing photometric data from poorly standardized catalogs carries with it inherent analysis uncertainties. In the X-ray study of \citet[][ see \S \ref{XMM}]{Pillitteri2004}, they use {\sc gsc-ii} $B$ and $R$ magnitudes transformed onto a version of the standard $B-V$ color system. If one compares these $B-V$ colors with those of stars in common in the J09 standardized photometric dataset for Blanco~1, a quality control assessment of the {\sc gsc-ii} transformation process can be performed. Plotted in the third panel of Fig.\ \ref{fig2} are the results of such a comparison, where a clear systematic offset between the two color systems is apparent. The Pillitteri et al. transformed $B-V$ colors are in fact systematically bluer, by $0.07$ magnitudes, than the J09 colors. Moreover, there is also considerable dispersion about the mean offset between the two color systems at the $10\%$ level ($1\sigma=0.12$). These absolute color calibration problems propagate throughout membership determination through isochrone fitting, as well as any calculations of bolometric luminosity for cataloged cluster members (e.g. determining $L_{x}/L_{bol}$). For the purpose of this paper, we employ only the $BVI_{c}$ data from the J09 survey.

\subsection{Proper Motions}\label{pm}
The recent astrometric study \citep{Platais2009} produced a new proper motion catalog for the central 8 deg$^2$ of Blanco~1. An astrometric solution was deduced from a total of 32 sets of photographic and {\sc ccd} observations with a time base-line of 40 years, ending in 2007 September. Proper motions and positions were calculated using a variant of the central plate-overlap method \citep[e.g.][]{Herbig1981} and the UCAC2 catalog \citep{Zacharias2004} as a reference frame. The precision of the proper motions, for stars with optimal image properties, is 0.3 mas~yr$^{-1}$. The final catalog contains 6300 objects down to $V$$\sim$17, among which, more than 3600 have proper-motion precisions better than 2 mas~yr$^{-1}$.

The formal proper-motion membership probabilities, $P_{\mu}$, were calculated using the probability definition formulated by \citet{Vasilevskis1958}: $P_{\mu} = \Phi_{c} / \Phi_{c} + \Phi_{f}$, where $\Phi_{c}$ is the distribution of cluster stars and $\Phi_{f}$ is the distribution of field star proper motions. The distributions of field and cluster stars in the area of Blanco~1 are derived using the so-called local sample method \citep{Platais2007}. The separation between the cluster and field is convincing for the entire magnitude range.  A total of 247 stars have their $P_{\mu}$ greater than 0\%. A full description of the reduction and analysis of these new astrometric data is given in \citet{Platais2009}.

Our new proper-motion membership probabilities can be used to scrutinize a list of astrometric Blanco~1 members given in Table~A.1  of \citet{Pillitteri2003}. Considering only those stars with $V$ brighter than $\sim$17, we identified 72 out of 93 objects on this list as cluster members based on new proper motions. Among the common stars, we find 17 stars with $P_{\mu}$=0\%. Apparently, the accuracy of {\sc gsc-ii} proper motions used by \citet{Pillitteri2003} is not adequate to efficiently eliminate field stars from their sample of cluster stars.

\subsection{X-Ray Observations}\label{XRO} 
\subsubsection{{\sc rosat}}\label{ROSAT} 
\citet[][ hereafter M99]{Micela1999a} report the results of two deep exposure ($\sim$70~ks) {\sc rosat hri}  pointings, bore-sighted on the central region of Blanco~1. The two adjacent $40\arcmin\times40\arcmin$ pointings were centered at {\sc ra}(2000): $00^{h} 02\fm8$; {\sc dec}(2000): $-30\arcdeg 00\arcmin$ (field~1) and {\sc ra}(2000): $00^{h} 05\fm6$; {\sc dec}(2000): $-30\arcdeg 06\arcmin$ (field 2). Identification of X-ray sources detected in these pointings was based upon a point-spread function ({\sc psf}) detection algorithm, with a source acceptance threshold chosen such that there is no more than one predicted false source detection per {\sc hri} image. This procedure yields a total of 132 X-ray sources from both fields in the 0.1 - 2.4 keV energy band. M99 utilize an X-ray error circle for significantly detected sources of $20\arcsec$. In order to derive fluxes for their {\sc hri} X-ray sources, M99 derives a count rate to flux conversion factor of $3.2 \times 10^{-11}$ erg cm$^{-2}$ cnt$^{-1}$ assuming a single temperature Raymond-Smith model for an optically thin plasma with a temperature of 1 keV and a hydrogen column density of $\log(N_{H})=20$.

In order to verify that identified X-ray sources are correlated with Blanco~1 cluster members, M99 employed the \citet{deEpstein1985} photometric membership list. They found 42 X-ray sources with optical counterparts, lying in X-ray positional errors circles, as well as having \citet{deEpstein1985} optical photometry consistent with being associated with the ``apparent'' Blanco~1 cluster main sequence. They additionally determined 41 X-ray flux upper limits for other likely cluster members, adjudged from de Epstein \& Epstein photometry. However, the de Epstein \& Epstein photographic photometry has considerable doubts as to its fidelity (see \S \ref{photscat}). These {\em must} act to introduce uncertainties into a photometric membership criterion, and thus, such optical/X-ray associations might include several spurious/suspect allocations of X-ray activity to uncertain cluster members. 

\subsubsection{{\sc xmm}-Newton}\label{XMM} 
\citet[][ hereafter P04]{Pillitteri2004} report the results arising from a deep exposure (50 ks) {\sc xmm}-Newton pointing of Blanco~1, centered on the coordinates of the field~1 pointing detailed in the M99 study. The observations were obtained with the {\sc epic} camera system, which has a field of view of $30\arcmin\times30\arcmin$. P04 used a {\sc psf} detection algorithm for source searching, which yielded a total of 190 X-ray sources detected in the 0.3 - 5.0 keV band. For {\sc xmm}-Newton, \citet{Jansen2001} states that the absolute location accuracy for the {\sc epic} instrument {\sc xmm}-Newton is uncertain up to 4$\arcsec$. Furthermore, P04 finds the internal precision for the {\sc epic} camera to be 2$\farcs$3 for the Blanco 1 field, thus giving a total positional uncertainty of 6$\farcs$3. Mirroring the M99 study, P04 set a source detection threshold such that no more than one spurious detection was predicted, with a key difference being that a positional error radius for X-ray sources of $13\arcsec$ was used. A total of 33 of the 190 {\sc xmm}-Newton X-ray sources are associated with the same optical counterpart in the M99 X-ray source list. P04 computed count rate to flux conversion factors for the 23 brightest X-ray sources from a detailed low-resolution spectral analysis using a grid of 2-T APEC models with photoelectric absorption. The spectra were found to be best modeled by a lower temperature component of 0.33 keV and an upper temperature component that varied typically from 0.8 to 1.5 keV and a hydrogen column density of $\log(N_{H}) = 20.5$. The count rate to flux conversions factors for these 23 stars were then averaged to get an overall conversion factor of $5.69 \times 10^{-12}$ erg cm$^{-2}$ cnt$^{-1}$ for the full {\sc xmm}-Newton X-ray dataset. 

P04 establishes cluster membership for Blanco~1 based on a \citet{Pillitteri2003} photometric and proper motion study. This earlier study uses a photometric selection based upon an $R$ versus $B-R$ {\sc cmd}, obtained from the second generation of the Guide Star Catalog ({\sc gsc-ii}) photometric data. Their analysis used a somewhat {\em ad hoc} by-eye selection, based on a region a few magnitudes wide around an assumed main-sequence locus. Furthermore, they refine their selection process by excluding photometric members with proper motion membership probabilities $p\leq0.8$. This methodology can lead to missing targets because proper motion membership probabilities can be dependent upon stellar magnitudes.

Armed with these membership constraints, P04 selected 93 stars as likely members of Blanco~1. Approximately $40\%$ of these stars ($36/93$) are optical counterparts to {\sc xmm}-Newton X-ray sources, including eight previously noted as nonmembers by the M99 study. Of the remaining 154 X-ray sources that were not determined as being Blanco~1 cluster members, 90 sources were found to have optical counterparts detailed in either the {\sc usno-b1}, {\sc gsc-ii}, or {\sc 2mass} catalogs. The remaining 64 (i.e. 154 total $-$90 matched sources) X-ray source detections could not be matched with optical counterparts. They are thus likely to be extra-Galactic background objects, which is hardly surprising given that Blanco~1 lies at high Galactic latitude ($b=-79\arcdeg$). In investigating the optical/X-ray relationships for Blanco~1, P04 used de Epstein \& Epstein $B-V$ for stars also found in the M99 study. As for the rest of the X-ray sources, they used the $B-V$ colors derived from the {\sc gsc-ii} $B-R$. Using these calculated $B-V$ colors which contain significant systematic scatter (see \S \ref{photscat}), introduces uncertainty in the optical/X-ray correlations derived in the P04 paper. 

\section{Revised Membership of X-ray Sources}\label{xrayfl} 
In order to determine the proper search radius for identifying optical counterparts in the J09 optical catalog to the M99 {\sc rosat} and P04 {\sc xmm}-Newton datasets, we employ the method outlined in \citet{Jeffries1997}. The procedure estimates the number of real versus spurious matches one should expect in a cross-correlation of an X-ray and optical catalog. This involves modeling the cumulative distribution of the closest match separations for the X-ray sources as the sum of two terms; the cumulative distribution of true correlations and the cumulative number of spurious sources which will increase with separation. For the {\sc rosat} dataset, we determine that a search radius of $16\arcsec$, which statistically should have $\sim$2/52 false counterpart matches, optimizes the number of counterpart matches while minimizing the expected spurious matches. The standard deviation of offsets for matched sources and optical counterparts for a $16\arcsec$ search radius is $4\farcs04$. In a similar fashion, we analyze the {\sc xmm}-Newton X-ray source matched with the J09 optical catalog. Our analysis showed that a search radius of $6\arcsec$ would maximize the number of matches while decreasing the expected spurious matches to $\sim$1/35. We find that the standard deviation of offsets for our {\sc xmm}-Newton matches is $1\farcs38$.

\begin{figure}[ht] \epsscale{0.75} \plotone{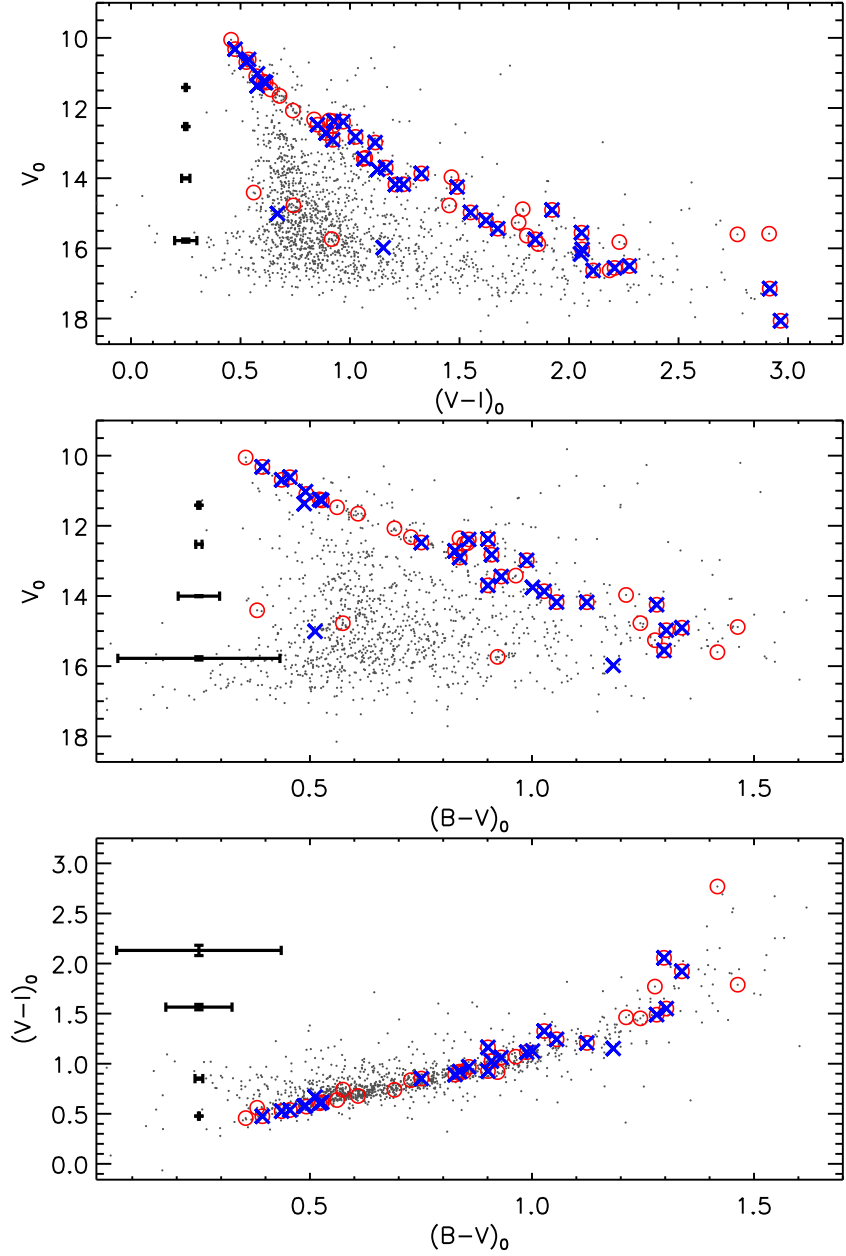} 
  \caption{
    \label{fig3}
    Same as Fig.\ \ref{fig1} however without the isochrones overplotted. In addition, we identify the optical counterparts to X-ray sources from {\sc rosat} 
    ({\it red open circles}) and {\sc xmm}-Newton ({\it blue crosses}). 
  } 
\end{figure}

\begin{figure}[ht] \epsscale{0.75} \plotone{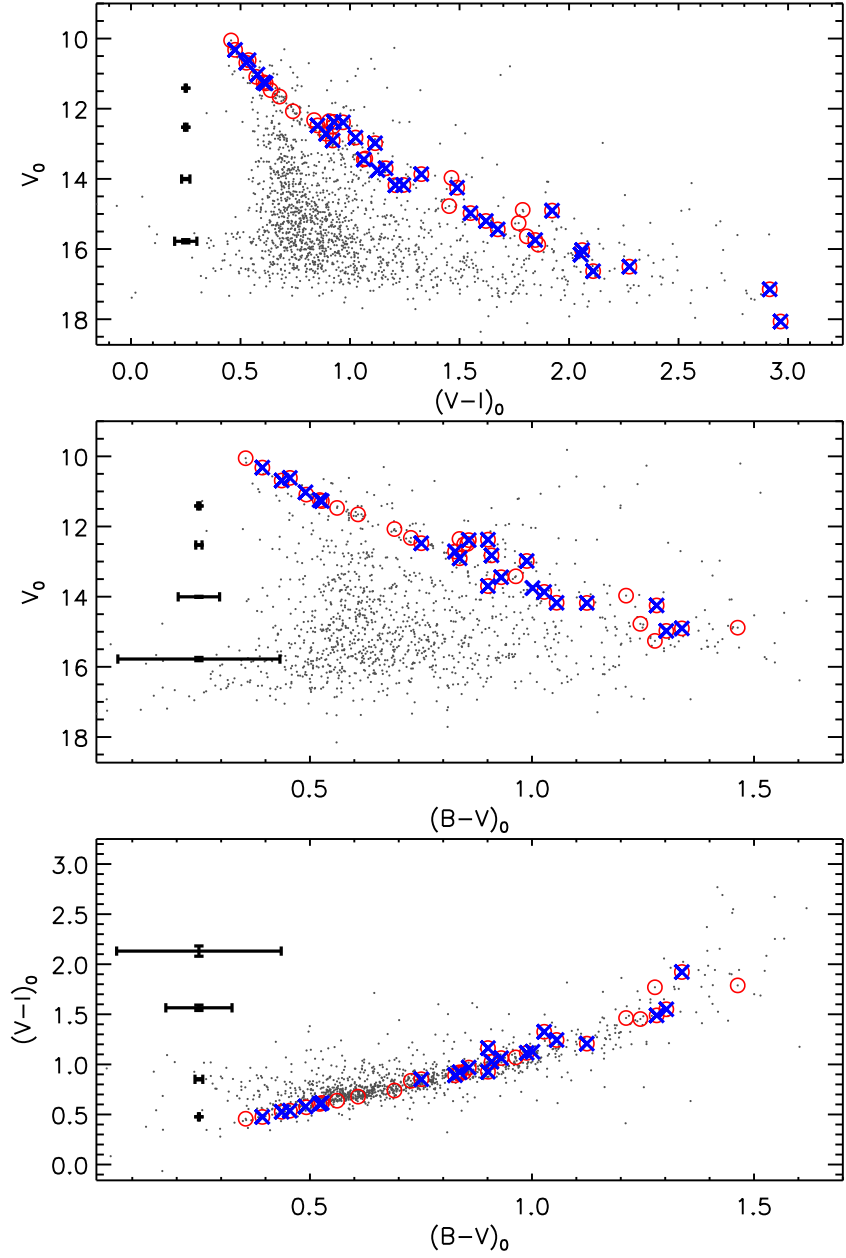} 
   \caption{
     \label{fig4}
     Same as Fig.\ \ref{fig3}, where {\sc rosat} ({\it red open circles}) and {\sc xmm}-Newton ({\it blue crosses}) are identified. 
     However, now only those objects are included that have proper motions consistent with the systemic motion of the cluster.
   } 
 \end{figure}

Using the search radii listed above ($16\arcsec$ for {\sc rosat} and $6\arcsec$ for {\sc xmm}-Newton), we matched optical counterparts in the J09 catalog to the X-ray sources published in the surveys of M99 and P04. The results of this matching are given in Table \ref{tab1}. To summarize our findings, in Fig.\ \ref{fig3} we plot the J09 optical catalog with the 52 {\sc rosat} and 35 {\sc xmm}-Newton X-ray sources with optical counterparts identified. Considering only those objects with proper motions consistent with the Blanco~1 membership, we find 41 X-ray sources that were previously identified as members of Blanco~1 by M99 and/or P04, including 24 stars observed by both telescopes. Interestingly, our new {\sc ccd} photometry and proper motion surveys have allowed us to identify an additional six X-ray sources which were incorrectly classified as nonmembers in the preceding X-ray surveys. Furthermore, based on the new proper motions, we reject three X-ray sources that M99 and P04 previously had identified as cluster members. These objects are likely active field stars or background active galaxies. The fact that these sources might be active galaxies is further supported by \citet{Richards2002} where they find that {\sc agn} in the {\sc sdss} database, with a wide range of redshifts, are clearly found with approximate colors $0.0 < B-V < 1.25$ and $0.0 < V-Ic < 1.5$ (using {\sc sdss} filter to $UBVI_{c}$ conversions of \citealt{Smith2002}). The unidentified sources in our X-ray dataset have optical counterparts with colors that fall directly within these {\sc agn} color ranges. In Fig.\ \ref{fig4}, we plot a {\sc cmd} marking only those X-ray sources that we have determined through proper motions to be cluster members. This demonstrates the power of precision photometric and proper motion data in defining a high fidelity cluster membership catalog. 

Of the total 47 X-ray sources we identify as being associated with cluster members, 26 have optical counterparts that are included in spectroscopic surveys of Blanco~1 \citep{Jeffries1999a,Mermilliod2008}. These surveys find that all 26 stars are identified as having radial velocities consistent with cluster membership. 

Moreover, we note that six optical counterparts in J09 are associated with two separate X-ray sources in M99 and P04. It appears that the {\sc xmm}-Newton sources that P04 identified as Blanco~1 nonmembers were not matched with the M99 {\sc rosat} X-ray dataset. Using the new proper motions, we subsequently find that three of these six stars are, in fact, cluster members (ZS44, BLX-16, BLX-42), while two (BLX-12 and BLX-15) fell below the proper motion survey's faintness limit. In Table \ref{tab2}, we list all six sources from M99 and P04 along with the single optical counterparts from J09. To be clear, the X-ray detections listed in Table \ref{tab2} are associated with the same optical counterpart, which M99 and P04 have identified as two separate X-ray sources. Apparently this is not the case. 

\subsection{X-Ray Luminosity and $L_{x}/L_{bol}$ ratio for Blanco~1 members}\label{xraylxlbol}
For cluster members which are optical counterparts to X-ray sources in Blanco~1, we derive their X-ray fluxes and luminosities using the published {\sc rosat} and {\sc xmm}-Newton X-ray count rates listed in M99 and P04. We convert count rates to fluxes applying the available conversion factors, $3.2\times10^{-11}$ erg cm$^{-2}$ cnt$^{-1}$ for {\sc rosat} and $5.69\times10^{-12}$ erg cm$^{-2}$ cnt$^{-1}$ for {\sc xmm}-Newton, and derived luminosities using the distance reported in J09, 240~pc. We propagate the uncertainties in flux from the published errors in the count rates. In the conversion from flux uncertainties to errors in X-ray luminosities, we use the distance error given in J09. For a selection of {\sc rosat} X-ray sources that did not have published uncertainties (see Table 2 in M99), we calculate their count rate errors using a linear interpolation of stars having both count rates and uncertainties in the M99 {\sc rosat} survey. 

\begin{figure}[ht] \epsscale{1.0} \plotone{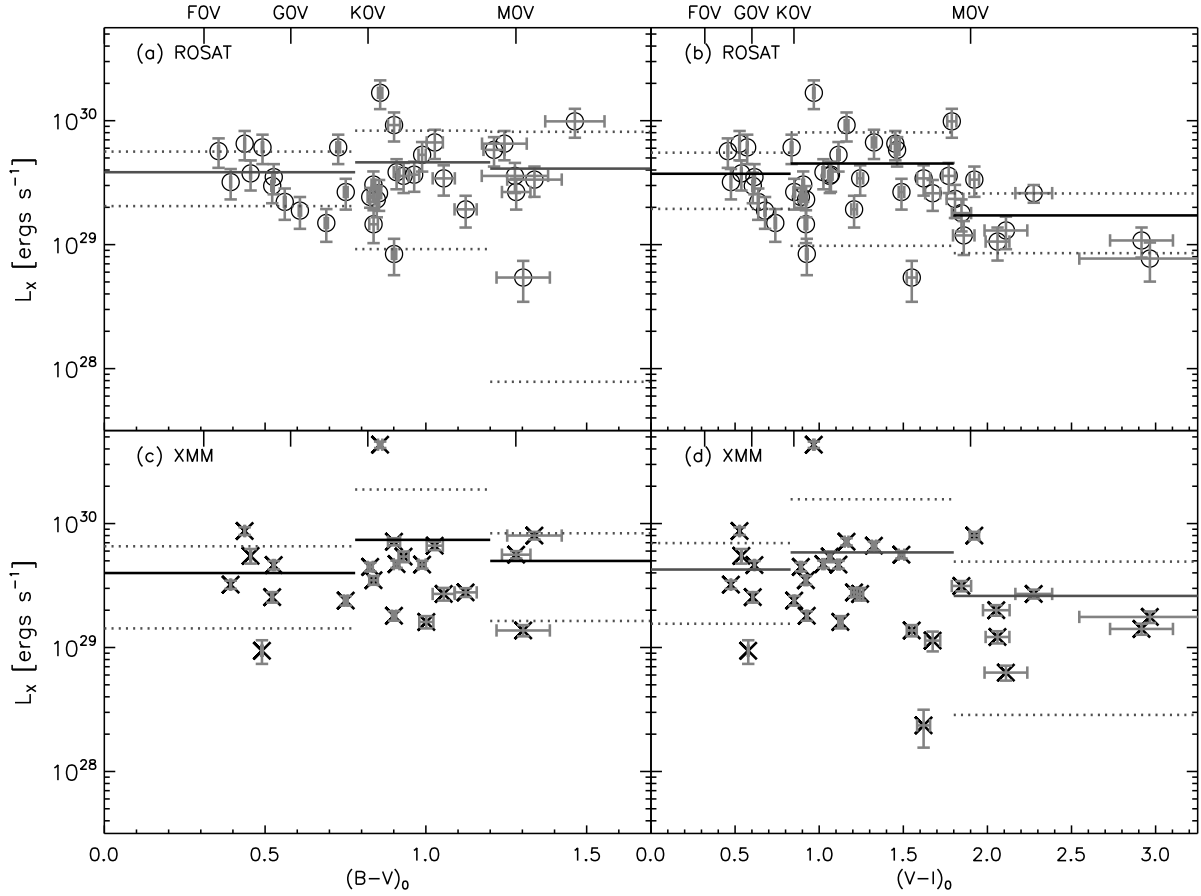} 
   \caption{
     \label{fig5}
     X-ray luminosity from {\sc rosat} ({\it open circles}) and {\sc xmm}-Newton ({\it crosses}) as a function of intrinsic $B-V$ and $V-I$ for Blanco~1 
     cluster members. The solid lines indicate the mean $L_{x}$ for F \& G, K, and M stars, where dotted lines represent $1\sigma$ values about these 
     means for each given spectral types. 
   } 
 \end{figure}

In Fig.\ \ref{fig5}, we show X-ray luminosity versus intrinsic $B-V$ and $V-I_{c}$ colors for both {\sc rosat} and {\sc xmm}-Newton Blanco~1 X-ray sources. We calculate intrinsic colors using $E(B-V) = 0.016$ and $E(V-I) = 0.02$ (see \S \ref{OP}).  Also plotted, and included in Table \ref{tab3}, are the mean X-ray luminosity and their 1$\sigma$ dispersions found for different spectral-types bins. These spectral bins are determined from color to spectral-type conversions defined in \citet{Kenyon1995}. Within the observed dispersion levels, the mean $L_{x}$ as a function of spectral type in the late F to early M star regime is constant. However, especially in the $V-I_{c}$ domain where lower mass stars are better represented, there is some evidence for a drop in X-ray luminosity. Empirically, observations of several open clusters support this conclusion (for instance, see \S \ref{xraycomp} and references therein.

\begin{figure}[ht] \epsscale{1.0} \plotone{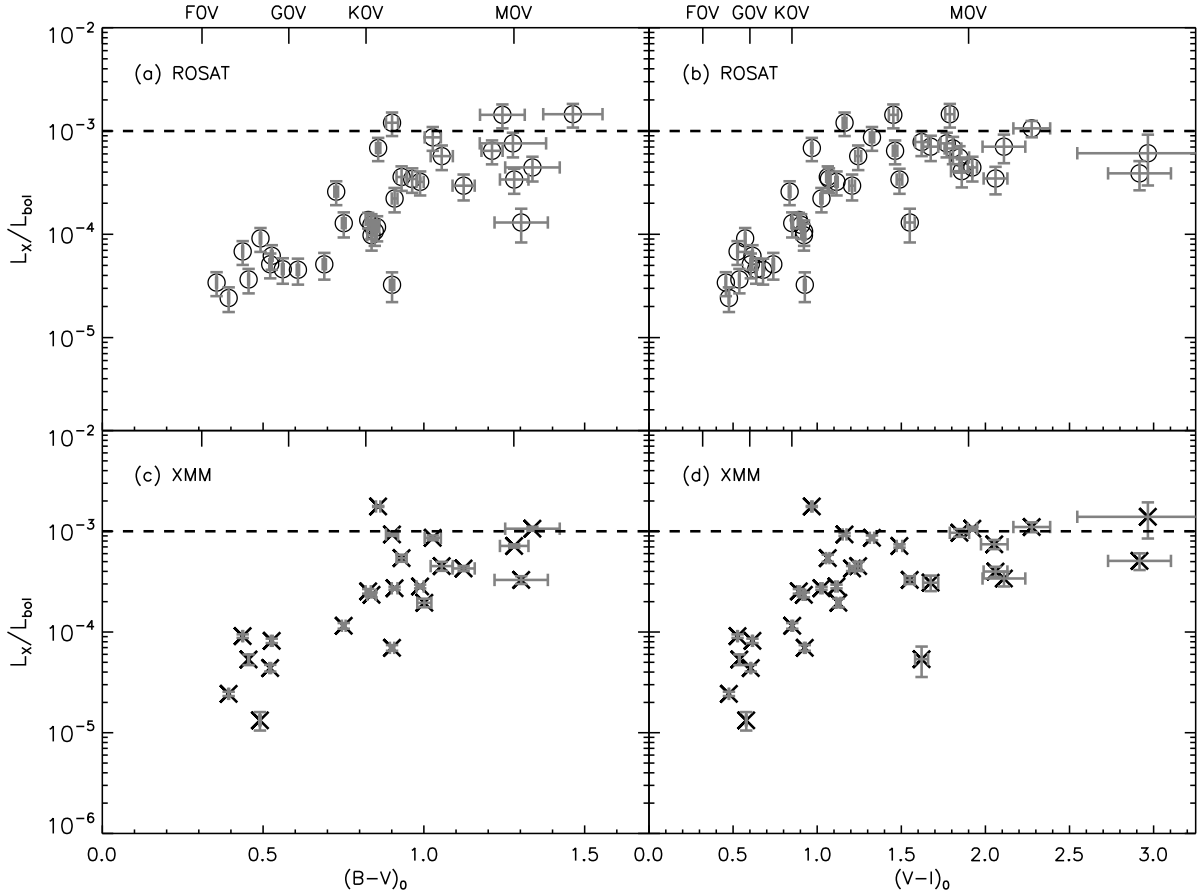}
   \caption{
     \label{fig6}
   Ratio of X-ray to bolometric luminosity plotted as a function of intrinsic color for {\sc rosat} ({\it open circles}) and {\sc xmm}-Newton 
   ({\it crosses}) sources in Blanco~1. The dashed line marks the level at which X-ray luminosity reaches 0.1 \% of the bolometric luminosity, which 
   is canonically known as X-ray saturation \citep[e.g.][]{Stauffer1994}. 
   }
 \end{figure}

We plot in Fig.\ \ref{fig6} the distance-independent ratio of X-ray to bolometric luminosity as a function of intrinsic $B-V$ and $V-I_{c}$ colors for Blanco~1 stars having X-ray detections. We note that in Fig.\ \ref{fig6}, as well as in Fig.\ \ref{fig5}, the $V-I_{c}$ color should be preferred for the reddest stars because for values greater than $\sim$1.4, $B-V$ becomes insensitive to changes in stellar mass. In our calculation of the bolometric luminosities, we use the bolometric corrections listed in \citet{Johnson1966} for stars with $V-I < 1.6$ and the formalism given in \citet{Monet1992} for stars with $V-I > 1.6$. A clear increase in $L_{x}/L_{bol}$ is observed as one goes from F-G to mid-K spectral types in Blanco~1. For spectral types later than K5, at $V-I \approx 1.25$ there is a saturation limit at a $L_{x}/L_{bol}$ of $10^{-3}$, where the X-ray production becomes insensitive to changes in spectral type. The exact cause of this limit has yet to be determined, however several theories have been put forward including limitations on the field generation capacity of the stellar dynamo \citep{Gilman1983,Vilhu1987} and/or due to centrifugal forces on magnetic loops in rapidly rotating stars \citep{Jardine1999, James2000}.

\subsection{Short-Term X-ray Variability}\label{xrayshort}
Previous X-ray studies, M99 and \citet{Pillitteri2005}, have provided in-depth investigations of short-term X-ray variability in Blanco~1. Therefore, here we merely provide a summary of those findings. M99 identified four variable X-ray sources in their {\sc rosat} dataset (ZS38,ZS61,ZS75,ZS76). We identify all four of these objects as proper motion members of Blanco~1. In a follow-up variability study of the P04 {\sc xmm}-Newton dataset, \citet{Pillitteri2005} found 22 variable X-ray sources. For 9 of these 22 sources we find optical counterparts in the J09 optical catalog, as well as identify them as proper motion members of Blanco~1 (ZS45,ZS46,ZS61,ZS75,ZS76,ZS94,ZS95, and BLX-42). 

It has been suggested that X-ray flaring in stellar coronae has a causal relationship to X-ray emission saturation. We find three of the four {\sc rosat} sources and four or the nine {\sc xmm}-Newton sources showing variability have saturated levels of X-ray emission, where saturation is arbitrarily defined as $L_{x}/L_{bol}$ above $10^{-3.25}$. In addition, nine {\sc rosat} and/or {\sc xmm}-Newton sources have saturated X-ray emission levels (ZS35,ZS37,ZS40,ZS42,ZS43,ZS53,ZS71,ZS88 and, ZS115) but were not identified by M99 and/or \citet{Pillitteri2005} as having significant X-ray variability. Assuming that the X-ray variability observed is evidence for flaring events, these statistics do not suggest that there is an intrinsic correlation between X-ray flaring and saturated X-ray emission levels.

\subsection{Long-Term X-ray Variability: Comparison of {\sc rosat} and {\sc xmm}-Newton Data Sets}\label{xrayvar}
\begin{figure}[ht] \epsscale{1.0} \plotone{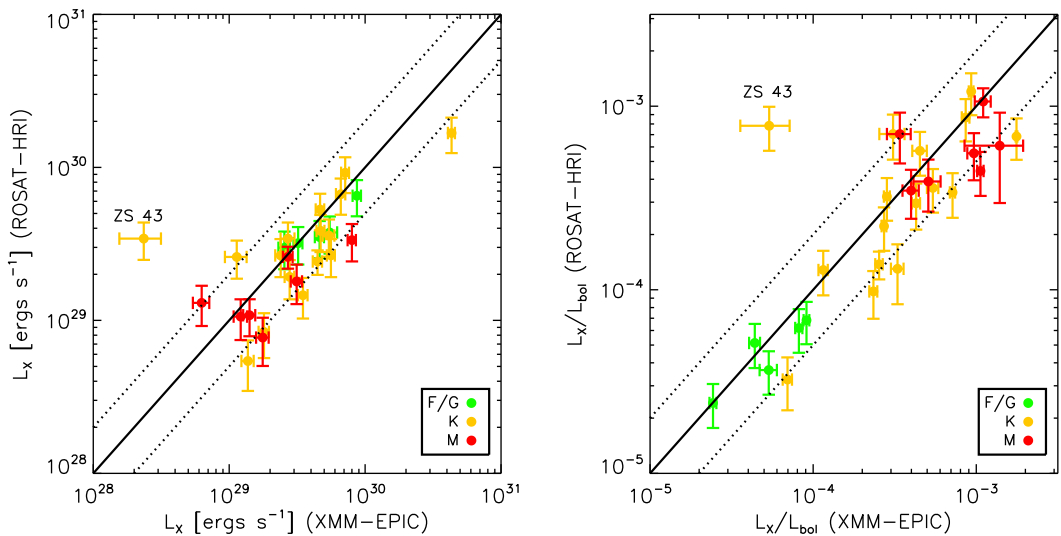}
   \caption{
     \label{fig7}
     Comparison of X-ray luminosity ({\it left}) and X-ray to bolometric luminosity ({\it right}) for the 28 cluster members observed with both {\sc rosat} and 
     {\sc xmm}-Newton, separated by $\sim$6 yr. The {\it solid} lines marks equality between measurements, whereas variations by factors of 0.5 and 2.0 
     ({\it dotted lines}) are also shown. We also color code these stars according to the different spectral ranges as defined by the colors given in 
     \citet{Kenyon1995}. One star (ZS43), that shows significant discordancy between the measurement systems, is identified. 
   } 
 \end{figure}

We cross-correlated the source positions listed in both surveys with a match radius of up to 16$\arcsec$, as derived by the positional uncertainties in the {\sc rosat} data. This search yields 28 matches for photometrically determined cluster members. For matched sources, we plot in Figs.\ \ref{fig7} the X-ray luminosity and the $L_{x}/L_{bol}$ values for both surveys along with lines representing equality and variations by factors of 0.5 and 2.

First, we note that the lack of any statistically significant systematic offset between the {\sc rosat} and {\sc xmm}-Newton X-ray luminosities/$L_{x}/L_{bol}$ values suggest that, even though the two datasets were observed over different energy bands, the conversion factors used to convert counts to X-ray fluxes were modeled correctly by M99 and P04. The reason for this agreement in X-ray flux over the two different energy bands is primarily due to the fact that the peak intensity of X-rays from Sun-like stars is around 1 keV \citep{Gudel2004}. Furthermore, this agreement provides assurance that our following analysis of X-ray emission from Blanco~1 stars does not suffer from instrumental systematics in the two X-ray datasets.

In order to investigate any possible long-term X-ray variability in Blanco~1, we look at the cluster members with X-ray sources in both {\sc rosat} and {\sc xmm}-Newton surveys. A time-span of $\sim$6 years separates these two surveys. These results suggest that the majority of the Blanco~1 cluster members have not undergone significant long-term X-ray variability. The source ZS43 shows a change in X-ray flux greater than a factor of 2, which is very likely due to the source confusion (see below). The apparent lack of long-term X-ray variability in Blanco~1 F-M dwarfs is in agreement with the {\sc rosat} and {\sc xmm}-Newton $L_{x}$ comparison made in \citet{Pillitteri2005}, as well as variability studies in other open clusters \citep[e.g. the Pleiades][]{Marino2003}. We note that for ZS76, a known X-ray flaring star \citep{Pillitteri2005}, we have adopted the count-rate of the star during its quiescent state.

As stated above, the {\sc rosat} X-ray flux measurement of ZS43 is significantly different than that of the {\sc xmm}-Newton measurements. This discrepancy is clearly seen in Fig.\ \ref{fig7}. A close inspection of the field near ZS43 reveals that another X-ray source, ZS42, lies only $11\farcs6$ away. The close proximity of these sources may lead to near-neighbor source confusion in the X-ray data, and therefore cause two-fold systematic offsets in the {\sc rosat} and {\sc xmm}-Newton X-ray fluxes for ZS43. First, due to the large {\sc psf} of the {\sc rosat} telescope ($>$16$\arcsec$), the counts for ZS43 in M99 likely includes flux from ZS42, and therefore the X-ray flux would be overestimated. Second, the {\sc xmm}-Newton {\sc psf} is smaller than {\sc rosat} and therefore the background level likely includes counts from ZS42 causing the {\sc xmm}-Newton X-ray flux to be underestimated. A combination of these two effects could explain the observed offset to the top left of this data point in Fig.\ \ref{fig7}.

\section{X-Ray Production in Blanco~1}\label{discusion}
\subsection{X-ray Activity along the Main Sequence}\label{xrayms}
\begin{figure}[ht] \epsscale{1.0} \plotone{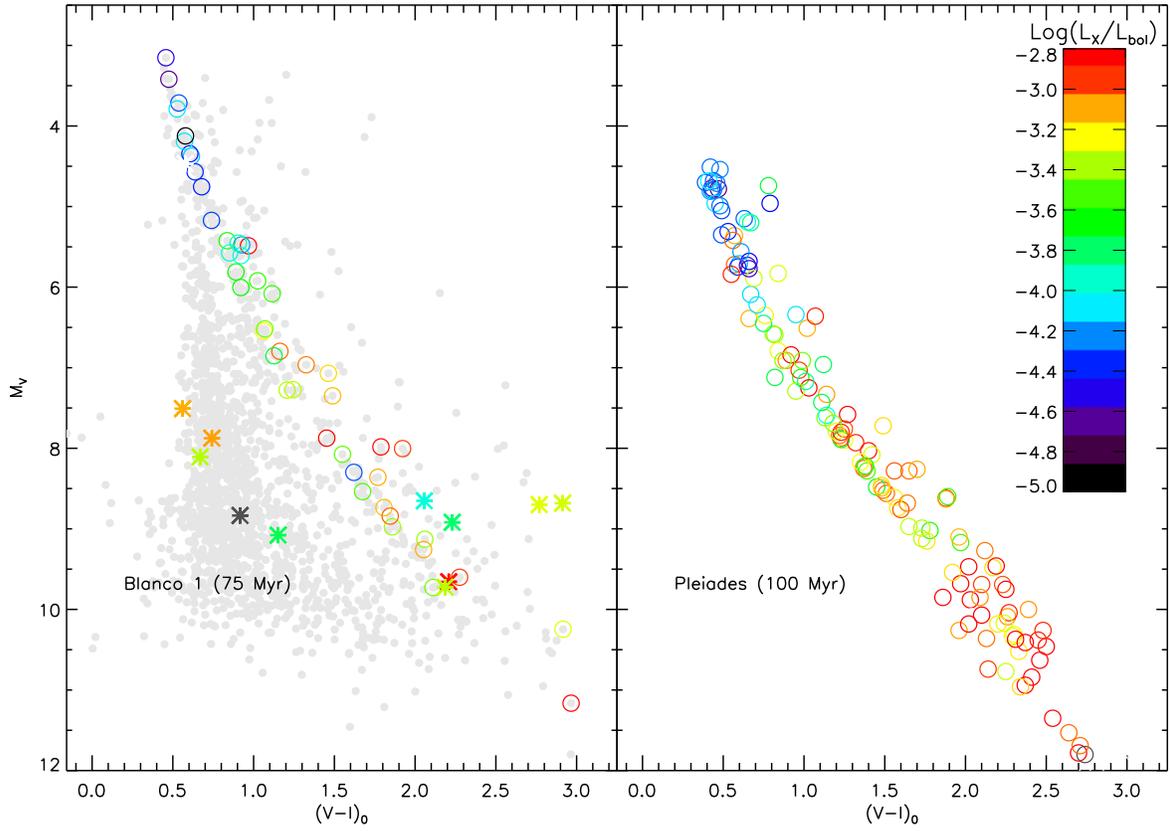}
   \caption{
     \label{fig8}
     Comparison of {\sc cmd} is shown for Blanco~1 {\it (left)} and Pleiades {\it (right)} open clusters, with optical counterparts to X-ray sources identified. 
     The magnitude of the ratio of X-ray to bolometric luminosity, given by the color coding, is shown on the right-hand panel. To derive the absolute magnitude 
     for the two clusters we use 240 (J09) and 133 \citep{Soderblom2005} parsecs for Blanco~1 and the Pleiades, respectively. For the Blanco~1 cluster, 
     nonmembers are noted with asterisks, and the full photometric catalog is shown in gray.
   }
 \end{figure}

In the left panel of Fig. \ref{fig8}, the $M_{V}$, $V-I$ {\sc cmd} for Blanco~1 is displayed showing the identified optical counterparts to X-ray sources and color-coded according to their magnitude of $L_{x}/L_{bol}$. We use the distance from J09 (240 parsecs) to derive the absolute magnitude for Blanco 1. We find X-ray counterparts to 47 optically-identified cluster members extending from early-F to mid-M spectral types. In Blanco~1, a general trend of increasing $L_{x}/L_{bol}$ with decreasing mass is seen along the main sequence. Although the {\sc psf} of {\sc rosat} and {\sc xmm}-Newton does not allow for observation of individual stars in known binary systems in Blanco~1, we do observe that photometric binaries of a given mass appear to be more X-ray luminous than their single star counterparts lying on the main sequence. This phenomenon is not unique to Blanco~1, with observations showing that binaries are typically over-luminous in X-rays when compared to single stars \citep[e.g.][]{Pye1994,Stern1995,Makarov2002}.  

\subsection{Activity-Age Relationship}\label{xraycomp}
It is known that there exist a strong causal relationship between X-ray emission and rotation rate in solar-type stars \citep{Pallavicini1981,Stauffer1994}. The underlining cause of this phenomenon can be understood in terms of greater dynamo-induced magnetic field production with increasing rotation rate \citep{Wilson1966,Kraft1967}. There is, however, an age effect to be considered. This is because, as solar-type stars age on the main sequence, they are capable of losing angular momentum through a magnetically channeled stellar wind \citep[e.g.][]{WeberDavis1967,Mestel1968,Kawaler1988,Barnes2003a}. Therefore, as stars become older their surface rotation rate decreases, resulting in an associated reduction in dynamo-induced magnetic field production. This is the so-called age-rotation-activity paradigm. This generalized scenario is observed in young open clusters and field stars as decay of magnetic activity and rotation in solar-type stars proportionally to inverse square root of their age ($t$) \citep{Skumanich1972}.

\begin{figure}[ht] \epsscale{1.0} \plotone{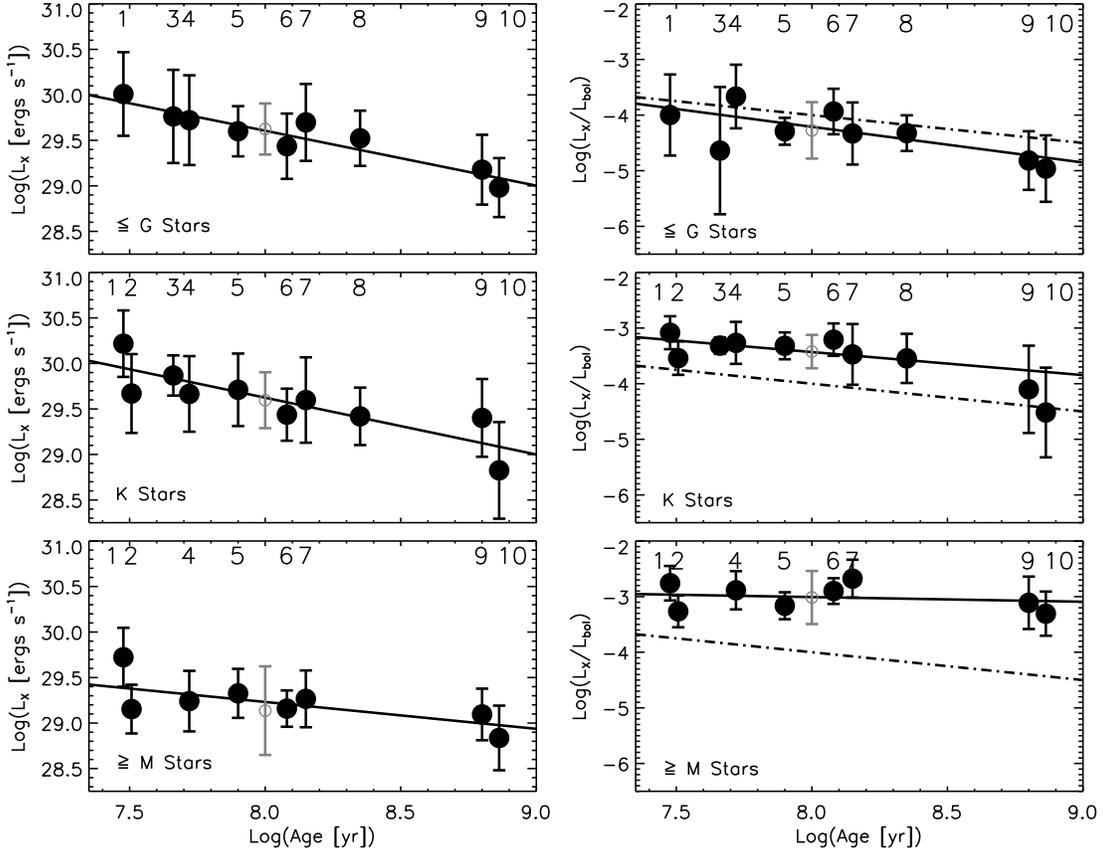} 
   \caption{
     \label{fig9}
     Evolution of $L_{x}$ {\it (left)} and $L{x}/L_{bol}$ {\it (right)} as a function of age is shown for these various open clusters detailed in Table 
     \ref{tab4}, including our new Blanco~1 measurements ({\it 5}) as well as those derived by P04 {\it (light-font, open circle)}. Reference numbers (1-10) 
     from Table 4 are shown in the upper part of each panel. Error bars represent the $1\sigma$ scatter for the given spectral ranges for each cluster. The 
     solid lines indicate linear, least-squares fits to the cluster data. In the right panel, the dash-dotted lines indicate a Skumanich-like 
     (i.e. $L_{x}/L_{bol} \sim Age^{-\frac{1}{2}}$) decay in X-ray emission. The initial conditions for the Skumanich functions are $L_{x}/L_{bol} = 
     10^{-3}$ at 1 Myr.
   }
 \end{figure}

In an effort to understand the relationship between activity and age in open clusters, we explore the mean X-ray luminosities of Blanco~1 stars in relation to several well studied open clusters of various ages. The results of this analysis are displayed in Fig.\ \ref{fig9} {\it (left panel)} and in Table \ref{tab4}. Two trends in the data are apparent. First, the mean X-ray luminosity of the F/G, K, and M stars decrease as a function of age. Our linear, least-squares fits to the data give a time dependence on X-ray luminosity of $L_{x}\varpropto$ $t^{-0.60\pm0.01}$, $t^{-0.62\pm0.27}$, $t^{-0.30\pm0.21}$ for spectral-type ranges of F/G, K, and M, respectively. This can be understood as stars having a less efficient dynamo with age because of stellar spin down. Second, the rate of decay is reduced in the M stars compared to the G and K stars in the surveyed clusters. This characteristic of the data is suggestive of longer spin down time scales for the lowest mass stars.

Similarly, we show in Fig.\ \ref{fig9} {\it (right panel)}, and in Table \ref{tab4}, the same cluster dataset as in Fig.\ \ref{fig9} {\it (left panel)}, this time substituting the mean $L_{x}/L_{bol}$ for mean $L_{x}$. We also compare a linear fit of the data to Skumanich-type spin down function (i.e. $L_{x}/L_{bol} \varpropto$ $t^{-\frac{1}{2}}$). For the Skumanich relation, we assume that on average all spectral types have a saturated X-ray level at an age of $\sim$1~Myr. This assumption is consistent with observations of very young open clusters \citep[e.g.][]{Feigelson2002,Stassun2004}. We find from our best-fit, linear trends that $L_{x}/L_{bol}$ is proportional to $t^{-0.64\pm0.41}$, $t^{-0.34\pm0.32}$, $t^{-0.08\pm0.26}$ for spectral-type ranges of F/G, K, and M, respectively. 

One can see some features in Fig.\ \ref{fig9} {\it (right panel)} that give insight into the activity-age relationship in solar-type stars. Formally, in all three mass regimes $L_{x}/L_{bol}$ decreases as a function of age. Moreover, the rate of decay of the $L_{x}/L_{bol}$ with age is mass dependent. The earliest spectral type stars in this sample, F- and G-type stars, appear to have $L_{x}/L_{bol}$ values which decay in a Skumanich-like manner. These higher mass stars are almost exclusively less X-ray active than the saturation level. K-type stars, however, follow a slower decay law of $\sim t^{-1/3}$. Looking more closely, these stars appear to have at or near saturated levels of X-ray emission for ages less than 100 Myr. In older clusters, K stars exhibit reduced magnitudes of X-ray emission due to their increased level of spin down, compared to the their younger ($<100$ Myr) counterparts. The M stars have an $L_{x}/L_{bol}$ evolution with age that is clearly non-Skumanich, and in fact, their magnitudes barely decay at all over the first 1 Gyr of their lives (although, see below). We note that our new Blanco~1 results, in $L_{x}$ and $L_{x}/L_{bol}$, are consistent with the X-ray/age dataset for both younger and older open clusters in the first $10^{9}$ years of stellar evolution.

We must caution the reader that by stating M dwarfs are only observed at or near saturated levels of X-ray emission, we are not implying that all young, low-mass stars ($<1$Gyr, M spectral types) have $L_{x}/L_{bol}$ magnitudes of 10$^{-3}$. Due to the limiting sensitivities of the {\sc rosat} and {\sc xmm}-Newton telescopes, the completeness level for X-ray surveys of open clusters decreases with decreasing mass. Therefore, the only X-ray sources observed at the lowest masses are the brightest X-ray sources, i.e. those with saturated levels of X-ray emission. In fact, one can observe this in the Hyades where, due to its proximity to the Sun, a near complete stellar population for the cluster is known down to a very low mass \citep{Reid1993,Bouvier2008}. In the {\sc rosat} study of the Hyades by \citet{Stern1995}, they observed X-ray emission from only 30\% of the known M dwarfs in the cluster, as compared to 90\% of X-ray activity in the known Hyades G stars. Thus, with the extant X-ray datasets for M dwarfs in young, $<1$Gyr, open clusters, we are unable to discriminate between the two scenarios of saturated levels of X-ray emission or sample incompleteness.

\begin{figure}[ht] \epsscale{1.00} \plotone{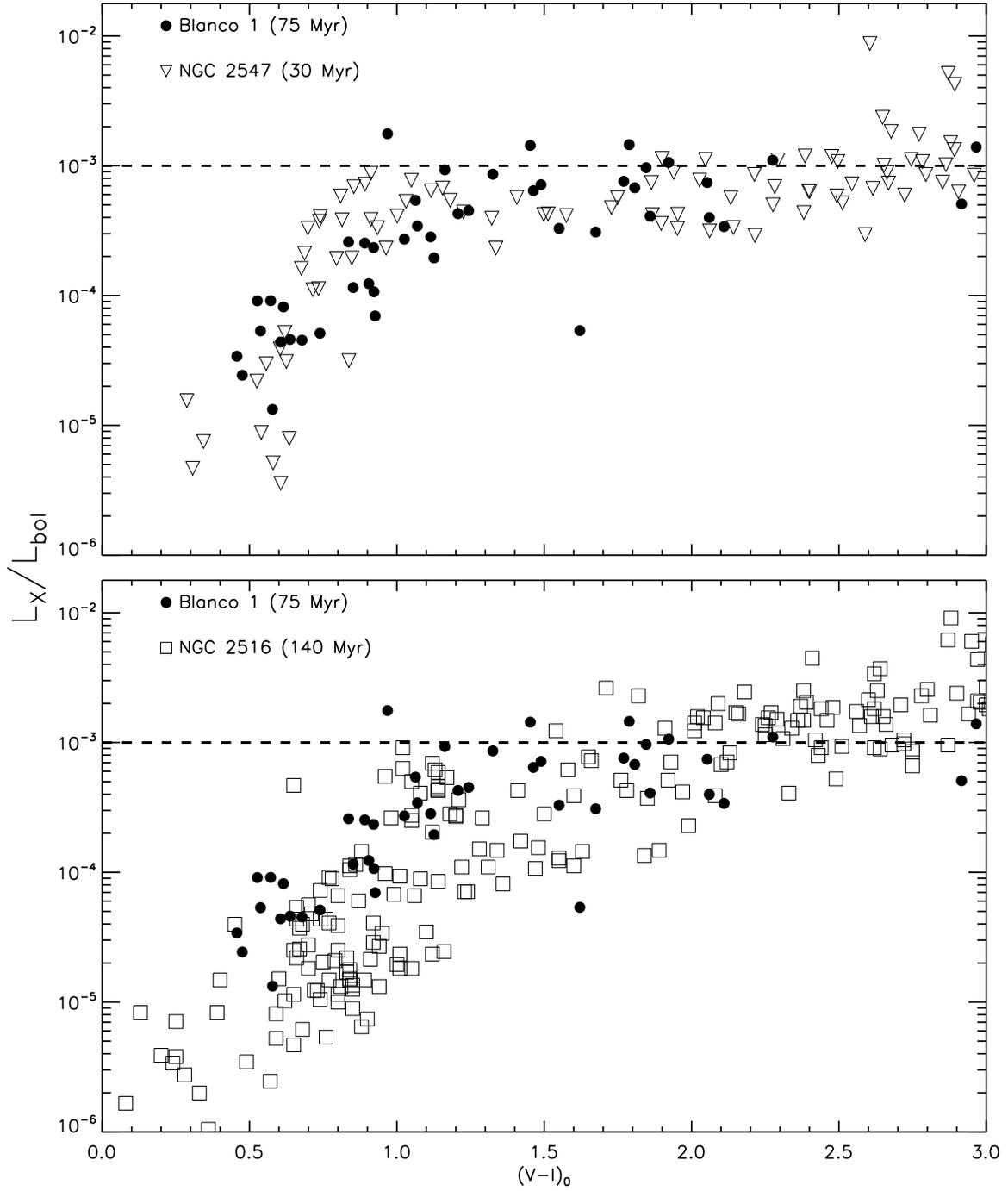}
   \caption{
     \label{fig10}
     $L_{x}/L_{bol}$ distributions for open clusters NGC~2547 ({\it triangles}) and NGC~2516 ({\it squares}) with the Blanco~1 distribution ({\it solid points}) 
     overplotted. The dashed line denotes a $L_{x}/L_{bol}$ of 10$^{-3}$, commonly referred to as the saturation point of X-ray emission. 
   }
 \end{figure}

Cognizant of some limitations in the existing X-ray datasets, let us continue.  In Fig.\ \ref{fig10}, the $L_{x}/L_{bol}$ distributions of two other well studied open clusters, NGC~2547 (age $\sim$ 30 Myr) and NGC~2516 (age $\sim$ 140 Myr), are directly compared to that of Blanco~1. The data for NGC~2547 and NGC~2516 are taken from \citet{Jeffries2006} and \citet{Pillitteri2006}, respectively. In NGC~2547, we generally see stars exhibiting saturated levels of X-ray emission at an earlier spectral type when compared to Blanco~1, in accordance with age-rotation-activity paradigm expectations. Thus, one would expect to still see rapidly rotating, higher mass stars that exhibit saturated levels of X-rays. In both clusters, almost all stars with $V-I_{c} \ge 1.25$ have saturated levels of X-ray emission. Therefore, in this mass regime, their ages and angular momentum as judged by X-ray emission are indistinguishable.

A comparison of the X-ray properties in Blanco~1 and NGC~2516 is somewhat less straightforward due to the considerable scatter in the $L_{x}/L_{bol}$ values for NGC~2516 stars at all masses. For the nonsaturated regime, the majority of Blanco~1 stars lie at or above the $L_{x}/L_{bol}$ distribution for NGC~2516. Moreover, while for Blanco~1 we judge by eye that saturation sets in at $V-I_{c} = 1.25\pm0.02$\footnote{Uncertainty determined from an average $V-I_{c}$ error for stars along the Blanco~1 main sequence with $V-I_{c}\sim$1.25.}, in NGC~2516 the nexus appears at a redder intrinsic color, $V-I_{c} = 1.5$. This finding is in agreement with the age-rotation-activity paradigm, that is, for the older NGC~2516, we expect to observe the point where stars exhibit saturated X-ray emission levels at a redder color (i.e. lower mass) when compared to the younger Blanco~1 open cluster. Finally, for the most part stars in the saturated regime of both Blanco~1 and NGC~2516 are indistinguishable in $L_{x}/L_{bol}$ versus intrinsic $V-I_{c}$ space. As we state above, this is most probably a result of flux sensitivity limits for the X-ray studies of these clusters, although the considerable scatter in the X-ray distribution of NGC~2516 clouds the issue somewhat. 

\subsection{Comparison to the Pleiades}\label{pleicomp}
\begin{figure}[ht] \epsscale{1.0} \plotone{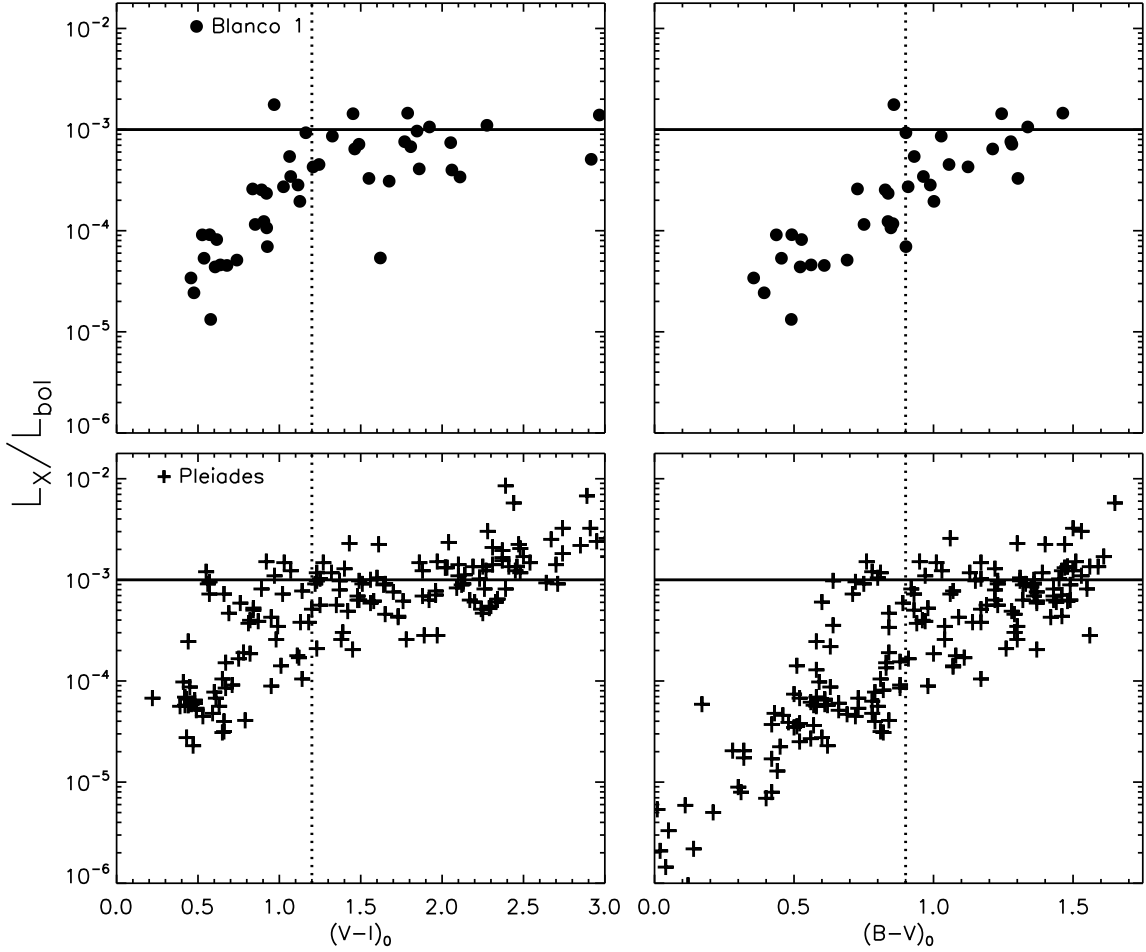}
   \caption{
     \label{fig11}
     A direct comparison of the X-ray to bolometric luminosity distributions for the Blanco~1 ({\it top}) and Pleiades ({\it bottom}) open clusters is plotted in 
     intrinsic $V-I_{c}$ ({\it left}) and $B-V$ ({\it right}) space, respectively. The Pleiades optical and X-ray data were taken from \citet{Stauffer1994,Micela1999b}
     and references therein. The solid lines represent an observed saturated level of $L_{x}/L_{bol} = 10^{-3}$. The dotted lines represent a by-eye determination of 
     the onset of saturation in Blanco~1. 
   }
 \end{figure}

Blanco~1 is oftentimes compared with the Pleiades cluster due to its similar age and metalicity (see \S \ref{Intro}). Under the umbrella of the activity-rotation-age paradigm, one would expect that the X-ray properties of these two clusters should be similar. In Fig.\ \ref{fig11}, we plot the $L_{x}/L_{bol}$ versus photometric color distributions for both clusters. The Pleiades optical and X-ray photometry is taken from the {\sc rosat} studies of \citet{Stauffer1994,Micela1999b} and references cited therein. The optical photometry for the Pleiades used by these authors is photoelectric where such data are available, and photographic otherwise. We notice in both clusters there are two clear distributions. First, there is an increasing level of $L_{x}/L_{bol}$ values (more active stars) as mass decreases for bluer, higher mass stars. Second, there is a plateau-like X-ray saturation for all redder, lower mass stars.

However, there does appear to be a difference between the two clusters in the photometric color (i.e. mass) at which X-ray saturation sets in. This mass appears to be higher for the Pleiades, which by the age-activity relationship would indicate that the Pleiads have not spun down as much as their Blanco~1 counterparts, and therefore would appear to be younger. This finding does not agree with previous age measurements for the two clusters. For Blanco~1, fitting of the main-sequence gives 80 Myr (J09); for the Pleiades, main sequence fitting gives 100 Myr \citep{Meynet1993} and the lithium-depletion boundary age is measured to be 125 Myr \citep{Stauffer1998}.
  
Looking at the {\sc cmd} for the X-ray sources for Blanco~1 and the Pleiades (Fig.\ \ref{fig8}), we clearly see that there is significant scatter across the Pleiades main sequence when compared to Blanco~1. The X-ray selected, photometric members of Blanco~1 appear to trace a much tighter locus in $M_{V}$/$V-I_{c}$ space when compared to the Pleiades cluster. Assuming that each X-ray identified Pleiad is a {\em bona fide} member of the cluster, only three possibilities can explain this phenomenon. First, the photometric spread on the main sequence is real. Second, the quality of the photometry is insufficient to define a tight main-sequence locus. Finally, differential reddening across the cluster is artificially introducing photometric scatter in the reddening-free $M_{V}$/$(V-I_{c})_0$ plane. One or all of these factors that lead to this scatter may well be contributing to some of the discrepancies seen in the $L_{x}/L_{bol}$ versus color distributions for Blanco~1 and the Pleiades. We do note that there is a larger number of sources identified in the Pleiades studies; therefore, the probability increases for including nonmembers in these datasets which coincidentally satisfy the criteria used for selection of cluster membership.

Much of the possible confusion in this discussion of age and X-ray activity at the onset of X-ray saturation in the Blanco~1 and Pleiades clusters centers upon the late F to early K stars ($0.6 < B-V < 0.9$; $0.5 < V-I_{c} < 1.1$). The morphology of these stars in the $L_{x}/L_{bol}$ versus intrinsic color plane appears unusual (see Fig.\ \ref{fig11}), with a clump of apparently saturated stars lying considerably above (up to an order of magnitude) the general trend of increasing $L_{x}/L_{bol}$ versus intrinsic color (especially in $B-V$ space). At the present time, with the available data, we cannot adequately explain this phenomenon.

\section{Summary}\label{summ}
Being young (80 Myr), nearby (240~pc), and having a high Galactic latitude ($b=-79\arcdeg$), the open cluster Blanco~1 presents itself as a valuable laboratory in which to study early stellar evolution. Here, we present a new analysis of the optical/X-ray properties for stars in Blanco~1 using the two extant X-ray surveys (M99 and P04) and a recent standardized $BVI_{c}$ photometric catalog (J09); membership selection of this cluster is based on newly derived proper motions. We find optical counterparts to 47 X-ray sources in the cluster. We note that six of these sources were misidentified as cluster nonmembers by previous X-ray studies. In our analysis, we derive new $L_{x}$ and $L_{x}/L_{bol}$ values for cluster members and compare the distribution of these parameters to other well studied open clusters. We find that the X-ray properties of Blanco~1 stars are in general agreement with those predicted by the age-rotation-activity paradigm. However, there is a disagreement between the distribution of $L_{x}/L_{bol}$ as a function of $B-V$ and $V-I_{c}$ color for Blanco~1 and the similar age Pleiades open cluster. This may be the result of large scatter seen in the color-magnitude diagram of the Pleiades X-ray sources, although the shift of a $L_{x}/L_{bol}$ saturation onset toward higher masses in the Pleiades appears to be larger than this scatter would imply. We do not find any evidence for significant long-term X-ray variability in the Blanco~1 cluster members.

Existing X-ray datasets (M99,P04) for Blanco~1, especially for objects fainter than a $V$$\sim$12, suffer from their reliance on photographic photometry. The analysis of the X-ray properties of Blanco 1 that we present in this manuscript supersedes these earlier analyses because our new study is founded upon a wide-field, high-quality, homogeneous optical photometric dataset, which crucially, we demonstrate is of high internal self-consistency. This property of the accompanying $BVI_{c}$ photometry allows us to describe the X-ray characteristics of stars in Blanco~1 as a function of mass, without some of the ambiguities affecting the earlier studies. 

As evidence of the power of this new standardized photometric dataset, the X-ray detected, proper motion members of Blanco~1 trace out a tight, low-dispersion main sequence, whereas in contrast the Pleiades cluster shows a much higher level of photometric scatter. We also observe that the absolute level of X-ray emission, as given by $L_{x}/L_{bol}$, changes along the Blanco~1 main sequence, thus clearly showing that X-ray production in Blanco~1 is mass dependent.

Looking more globally at the connatural relationship between stellar mass, X-ray production, and age, we observe that X-ray production from the stars in Blanco~1 follows very distinct trends seen in other open clusters. Namely, X-ray emission decays as a function of age, and this decay is mass dependent. G-type stars of all ages have X-ray emission that decays with a Skumanich-like trend. However, as one looks at lower mass stars, X-ray emission become less of a function of stellar age. In fact, for the lowest mass (M-type) stars, there is no observable evidence for the reduction of X-ray emission during the first 1 Gyr of their lives, however, due to the limiting flux sensitivities of the {\sc rosat} and {\sc xmm}-Newton datasets, we are probably not able to detect M-dwarf stars with less than saturated levels of X-ray emission. 

\acknowledgments
We cordially thank J.-C. Mermilliod for his many years of service to the field of astronomy and to the study of open cluster, including his valuable contribution to this paper on Blanco~1. We recognize support for P.A.C. and D.J.J. from the National Science Foundation Career Grant AST-0349075 (P.I. Stassun, K.~G.). I. P. gratefully acknowledges support from the National Science Foundation through grant AST 04-06689 to Johns Hopkins University. Dr. Kelly~Holley-Bockelmann and an anonymous referee are thankfully acknowledged for input on this manuscript. We also gratefully acknowledge the staff at {\sc ctio} and the {\sc smarts} Consortium. The Cerro Tololo Inter-American Observatory, and the National Optical Astronomy Observatory, are operated by the Association of Universities for Research in Astronomy, under contract with the National Science Foundation. This research has made use of the {\sc simbad} database, operated at {\sc cds}, Strasbourg, France.

{\it Facilities:} \facility{ROSAT ()}, \facility{XMM ()}, \facility{CTIO:1.0m ()}

  \clearpage 
 
  \clearpage 

 \clearpage

  \clearpage 
 
  \clearpage 

  \clearpage 

  \clearpage 

  \clearpage 

  \clearpage 

  \clearpage

  \clearpage 

 \clearpage 

 \begin{deluxetable}{l c c c c c c c c c c c c c}
\tablecolumns{14}
\tablewidth{0pt}
\rotate
\tabletypesize{\scriptsize}
\tablecaption{Blanco~1 X-ray Source Catalog\label{tab1}}
\tablehead{
\colhead{} & \colhead{} & \colhead{} & \colhead{} & \colhead{} & \colhead{} & \multicolumn{2}{c}{{\sc rosat} [0.1-2.4 keV]} & \multicolumn{2}{c}{{\sc xmm-newton} [0.3-5.0 keV]} & \colhead{} & \colhead{} & \colhead{} & \colhead{} \\
\colhead{ID\tablenotemark{a}} & \colhead{RA\tablenotemark{b}} & \colhead{DEC\tablenotemark{b}} & \colhead{$V_{0}$\tablenotemark{c}} & \colhead{$(V-I_{c})_{0}$\tablenotemark{c}} & \colhead{$(B-V)_{0}$\tablenotemark{c}} & \colhead{$Log L_{x}$\tablenotemark{d}} & \colhead{$Log$} & \colhead{$Log L_{x}$\tablenotemark{d}} & \colhead{$Log$} & \colhead{$\mu_{\alpha}\cos{\delta}$\tablenotemark{e}} & \colhead{$\mu_{\delta}$\tablenotemark{e}} & \colhead{P$_{\mu}$\tablenotemark{e}} & \colhead{P.M.\tablenotemark{e}} \\
 & \colhead{[HH:MM:SS]} & \colhead{[DD:MM:SS]} & \colhead{} & \colhead{} & \colhead{} & \colhead{} & \colhead{$L_{x}/L_{bol}$} & \colhead{} & \colhead{$L_{x}/L_{bol}$} & \colhead{} & \colhead{} & \colhead{} & \colhead{Mem.}
}
\startdata
 ZS35                    &   00:01:39.85 & -30:04:38.4  &  14.78$\pm$0.022 &   1.45$\pm$0.027 &   1.24$\pm$0.070 & 29.81 & -2.84 & ...   &  ...   & 22.0 &   1.7 & 60 &  Y  \\
 ZS58                    &   00:01:46.46 & -29:46:38.7  &  12.32$\pm$0.003 &   0.84$\pm$0.005 &   0.73$\pm$0.006 & 29.78 & -3.59 & ...   &  ...   & 20.8 &   2.5 & 86 &  Y  \\
 ZS37                    &   00:01:53.42 & -30:06:12.9  &  15.74$\pm$0.052 &   1.85$\pm$0.057 &         ...      & 29.25 & -3.26 & 29.49 & -3.02  & 22.3 &   3.0 & 58 &  Y  \\
 ZS38                    &   00:01:54.44 & -30:07:41.8  &  13.86$\pm$0.010 &   1.33$\pm$0.013 &   1.03$\pm$0.025 & 29.82 & -3.06 & 29.82 & -3.07  & 22.0 &   2.4 & 86 &  Y  \\
 ZS40                    &   00:01:56.92 & -30:12:07.9  &  15.44$\pm$0.039 &   1.67$\pm$0.045 &         ...      & 29.41 & -3.15 & 29.05 & -3.51  & 22.7 &   2.6 & 54 &  Y  \\
 BLX-7                   &   00:02:00.80 & -29:59:17.4  &  12.71$\pm$0.004 &   0.89$\pm$0.006 &   0.83$\pm$0.008 & 29.38 & -3.86 & 29.65 & -3.60  & 21.5 &   1.9 & 80 &  Y  \\
 ZS43                    &   00:02:03.69 & -30:10:24.9  &  15.20$\pm$0.033 &   1.62$\pm$0.039 &         ...      & 29.53 & -3.11 & 28.37 & -4.27  & 22.8 &   4.0 & 54 &  Y  \\
 ZS42                    &   00:02:04.21 & -30:10:34.4  &  14.17$\pm$0.013 &   1.24$\pm$0.018 &   1.06$\pm$0.035 & 29.53 & -3.24 & 29.43 & -3.35  & 21.8 &   3.1 & 89 &  Y  \\
 BLX-12\tablenotemark{g} &   00:02:07.66 & -30:04:42.6  &  18.06$\pm$0.414 &   2.97$\pm$0.420 &    ...           & 28.88 & -3.22 & 29.24 & -2.86  & ...  &   ... & ... & NA \\
 ZS44\tablenotemark{g}   &   00:02:14.69 & -29:49:04.2  &  13.45$\pm$0.007 &   1.06$\pm$0.010 &   0.93$\pm$0.016 & 29.55 & -3.45 & 29.73 & -3.27  & 22.1 &   3.0 & 92 &  Y  \\
 ZS45                    &   00:02:18.55 & -29:51:08.5  &  12.91$\pm$0.006 &   0.92$\pm$0.008 &   0.84$\pm$0.013 & 29.16 & -4.01 & 29.54 & -3.63  & 22.1 &   2.0 & 84 &  Y  \\
 ZS46                    &   00:02:19.73 & -29:56:07.5  &  14.25$\pm$0.014 &   1.49$\pm$0.017 &   1.28$\pm$0.044 & 29.42 & -3.47 & 29.74 & -3.15  & 20.1 &   1.5 & 16 &  Y  \\
 ZS48                    &   00:02:21.63 & -30:08:21.6  &  10.69$\pm$0.001 &   0.53$\pm$0.002 &   0.44$\pm$0.001 & 29.81 & -4.17 & 29.93 & -4.04  & 20.7 &   3.3 & 90 &  Y  \\
 BLX-15\tablenotemark{g} &   00:02:22.90 & -30:02:53.2  &  17.15$\pm$0.184 &   2.92$\pm$0.187 &    ...           & 29.03 & -3.41 & 29.15 & -3.29  & ...  &   ... & ... & NA \\
 BLX-16\tablenotemark{g} &   00:02:23.63 & -29:50:39.8  &  16.03$\pm$0.066 &   2.06$\pm$0.071 &    ...           & 29.02 & -3.46 & 29.08 & -3.40  & 19.3 &   5.6 & 03 &  Y  \\
 ZS53                    &   00:02:24.26 & -30:09:09.0  &  16.16$\pm$0.074 &   2.05$\pm$0.079 &        ...       & ...   & ...   & 29.30 & -3.13  & 20.3 &  -0.7 & 01 &  Y  \\
 PMS04-94                &   00:02:25.48 & -29:59:17.6  &  15.98$\pm$0.065 &   1.15$\pm$0.087 &   1.18$\pm$0.195 & ...   &  ...  & 28.37 & -3.76  & 15.0 &  20.5 & 00 &  N  \\
 BLX-17                  &   00:02:25.89 & -29:52:39.2  &  16.56$\pm$0.105 &   2.21$\pm$0.111 &         ...      & 29.04 & -3.30 & 29.45 & -2.89  & 16.8 &  10.9 & 00 &  N  \\
 ZS54                    &   00:02:28.19 & -30:04:43.5  &  12.98$\pm$0.005 &   1.11$\pm$0.007 &   0.99$\pm$0.011 & 29.72 & -3.49 & 29.66 & -3.55  & 21.8 &   2.5 & 90 &  Y  \\
 ZS61                    &   00:02:34.83 & -30:05:25.5  &  13.70$\pm$0.008 &   1.16$\pm$0.011 &   0.90$\pm$0.019 & 29.96 & -2.92 & 29.85 & -3.03  & 21.3 &   2.1 & 80 &  Y  \\
 ZS62\tablenotemark{\dagger} &   00:02:35.46 & -30:07:02.0  &  12.48$\pm$0.003 &   0.85$\pm$0.005 &   0.75$\pm$0.006 & 29.42 & -3.89 & 29.37 & -3.94  & 21.2 &   2.9 & 93 &  Y  \\
 ZS60\tablenotemark{g}   &   00:02:41.79 & -29:58:53.2  &  15.55$\pm$0.043 &   2.06$\pm$0.046 &   1.30$\pm$0.147 & 29.10 & -3.57 & 28.70 & -3.97  &-17.8 & -12.0 & 00 &  N  \\
 PMS04-190               &   00:02:48.22 & -29:46:34.9  &  13.75$\pm$0.009 &   1.13$\pm$0.013 &   1.00$\pm$0.022 & ...   &  ...  & 29.20 & -3.71  & 20.9 &   3.2 & 86 &  Y  \\
 BLX-26                  &   00:02:51.52 & -29:54:49.4  &  16.63$\pm$0.119 &   2.11$\pm$0.127 &         ...      & 29.11 & -3.15 & 28.79 & -3.47  & 18.5 &   3.1 & 04 &  Y  \\
 ZS76                    &   00:02:56.38 & -30:04:44.8  &  12.39$\pm$0.003 &   0.97$\pm$0.004 &   0.86$\pm$0.007 & 30.22 & -3.16 & 30.63 & -2.75  & 21.6 &   3.2 & 95 &  Y  \\
 BLX-32                  &   00:02:59.65 & -29:52:52.2  &  15.82$\pm$0.055 &   2.23$\pm$0.058 &         ...      & 28.84 & -3.79 & ...   &  ...   &135.5 &  81.8 & 00 &  N  \\
 ZS75                    &   00:03:00.28 & -30:03:21.6  &  12.82$\pm$0.004 &   1.03$\pm$0.006 &   0.91$\pm$0.009 & 29.58 & -3.65 & 29.67 & -3.57  & 21.5 &   3.9 & 91 &  Y  \\
 BLX-34                  &   00:03:00.56 & -30:15:44.0  &  15.64$\pm$0.044 &   1.81$\pm$0.050 &         ...      & 29.36 & -3.17 & ...   &  ...   & 21.0 &   1.6 & 33 &  Y  \\
 ZS71                    &   00:03:02.95 & -29:47:44.1  &  14.90$\pm$0.025 &   1.92$\pm$0.028 &   1.34$\pm$0.085 & 29.52 & -3.35 & 29.90 & -2.97  & 21.2 &   3.6 & 80 &  Y  \\
 ZS88                    &   00:03:06.63 & -29:43:11.5  &  13.97$\pm$0.011 &   1.46$\pm$0.014 &   1.21$\pm$0.033 & 29.76 & -3.19 & ...   &  ...   & 21.1 &   5.0 & 45 &  Y  \\
 ZS83\tablenotemark{\ddagger} &   00:03:07.09 & -30:15:17.1  &  12.51$\pm$0.003 &   0.92$\pm$0.005 &   0.85$\pm$0.007 & 29.36 & -3.97 & ...   &  ...   & 21.9 &   2.6 & 94 &  Y  \\
 ZS84                    &   00:03:10.81 & -30:10:49.0  &  11.28$\pm$0.002 &   0.62$\pm$0.003 &   0.53$\pm$0.003 & 29.54 & -4.21 & 29.66 & -4.09  & 23.2 &   3.9 & 82 &  Y  \\
 ZS95                    &   00:03:16.49 & -29:58:47.4  &  12.38$\pm$0.003 &   0.93$\pm$0.005 &   0.90$\pm$0.007 & 28.92 & -4.49 & 29.25 & -4.16  & 20.1 &   2.7 & 60 &  Y  \\
 ZS91                    &   00:03:20.61 & -29:49:22.8  &  11.25$\pm$0.002 &   0.61$\pm$0.003 &   0.52$\pm$0.003 & 29.47 & -4.29 & 29.40 & -4.36  & 22.1 &   2.5 & 95 &  Y  \\
 ZS96\tablenotemark{\dagger} &   00:03:21.85 & -30:01:10.5  &  10.32$\pm$0.001 &   0.48$\pm$0.001 &   0.39$\pm$0.001 & 29.50 & -4.62 & 29.50 & -4.61  & 20.8 &   3.5 & 92 &  Y  \\
 BLX-42\tablenotemark{g} &   00:03:22.73 & -29:53:50.5  &  16.50$\pm$0.104 &   2.28$\pm$0.109 &    ...           & 29.41 & -2.97 & 29.43 & -2.96  & 23.0 &   7.4 & 02 &  Y  \\
 ZS94                    &   00:03:24.18 & -29:56:22.9  &  14.98$\pm$0.025 &   1.55$\pm$0.030 &   1.30$\pm$0.083 & 28.73 & -3.89 & 29.13 & -3.48  & 22.3 &   3.3 & 79 &  Y  \\
 ZS90                    &   00:03:24.39 & -29:48:49.4  &  10.62$\pm$0.001 &   0.54$\pm$0.002 &   0.45$\pm$0.001 & 29.57 & -4.44 & 29.73 & -4.27  & 20.7 &   3.5 & 90 &  Y  \\
 ZS93                    &   00:03:24.67 & -29:55:14.7  &  14.18$\pm$0.013 &   1.21$\pm$0.017 &   1.12$\pm$0.035 & 29.28 & -3.53 & 29.44 & -3.37  & 22.2 &   2.4 & 83 &  Y  \\
 ZS104                   &   00:03:31.89 & -29:43:04.8  &  10.05$\pm$0.001 &   0.46$\pm$0.001 &   0.36$\pm$0.001 & 29.75 & -4.47 & ...   &  ...   & 22.8 &   3.0 & 95 &  Y  \\
 PMS04-70                &   00:03:39.79 & -30:02:09.5  &  11.37$\pm$0.002 &   0.58$\pm$0.003 &   0.49$\pm$0.003 & ...   &  ...  & 28.59 & -5.12  & -4.9 &  -4.6 & 00 &  N  \\
 ZS107\tablenotemark{\dagger \dagger} &   00:03:50.17 & -30:03:55.7  &  11.02$\pm$0.001 &   0.58$\pm$0.002 &   0.49$\pm$0.002 & ...   & ...   & 28.95 & -4.88  & 22.0 &   2.9 & 97 &  Y  \\
 ZS115                   &   00:04:12.57 & -29:58:02.5  &  14.88$\pm$0.024 &   1.79$\pm$0.027 &   1.46$\pm$0.092 & 29.99 & -2.84 & ...   &  ...   & 21.7 &   4.3 & 70 &  Y  \\
 ZS134                   &   00:04:49.20 & -30:00:52.9  &  11.09$\pm$0.001 &   0.57$\pm$0.002 &   0.49$\pm$0.002 & 29.78 & -4.04 & ...   &  ...   & 23.4 &   5.0 & 29 &  Y  \\
 ZS138                   &   00:04:58.84 & -30:09:41.6  &  11.47$\pm$0.002 &   0.64$\pm$0.003 &   0.56$\pm$0.003 & 29.34 & -4.34 & ...   &  ...   & 22.4 &   2.6 & 95 &  Y  \\
 ZS142                   &   00:05:04.93 & -30:19:39.2  &  15.60$\pm$0.045 &   2.77$\pm$0.046 &   1.42$\pm$0.159 & 29.69 & -3.29 & ...   &  ...   & 99.1 &   1.2 & 00 &  N  \\
 BLX-62                  &   00:05:07.01 & -30:04:29.3  &  14.78$\pm$0.021 &   0.74$\pm$0.034 &   0.57$\pm$0.037 & 29.26 & -3.10 & ...   &  ...   &  8.7 &  -6.8 & 00 &  N  \\
 ZS144                   &   00:05:07.08 & -29:59:25.7  &  15.26$\pm$0.032 &   1.77$\pm$0.037 &   1.28$\pm$0.103 & 29.55 & -3.12 & ...   &  ...   & 21.8 &   5.0 & 33 &  Y  \\
 ZS148                   &   00:05:14.39 & -29:54:23.8  &  12.35$\pm$0.003 &   0.90$\pm$0.005 &   0.84$\pm$0.006 & 29.48 & -3.91 & ...   &  ...   & 22.8 &   4.0 & 87 &  Y  \\
 ZS154\tablenotemark{\ddagger} &   00:05:31.58 & -30:20:51.6  &  13.42$\pm$0.007 &   1.07$\pm$0.010 &   0.96$\pm$0.016 & 29.56 & -3.46 & ...   &  ...   & 22.0 &   2.2 & 84 &  Y  \\
 ZS165\tablenotemark{\dagger}  &   00:05:35.53 & -29:57:06.4  &  12.50$\pm$0.003 &         ...      &   0.85$\pm$0.007 & 29.41 & -3.93 & ...   &  ...   & 21.0 &   3.8 & 89 &  Y  \\
 ZS170                   &   00:05:54.72 & -30:06:25.8  &  12.07$\pm$0.003 &   0.74$\pm$0.004 &   0.69$\pm$0.005 & 29.17 & -4.29 & ...   &  ...   & 22.0 &   2.5 & 94 &  Y  \\
 BLX-79                  &   00:05:58.13 & -30:11:09.0  &  16.63$\pm$0.112 &   2.19$\pm$0.119 &         ...      & 28.94 & -3.36 & ...   &  ...   & 27.5 &  -5.8 & 00 &  N  \\
 ZS172                   &   00:06:04.29 & -30:02:11.9  &  15.88$\pm$0.058 &   1.86$\pm$0.064 &         ...      & 29.07 & -3.39 & ...   &  ...   & 21.3 &   3.2 & 49 &  Y  \\
 BLX-81                  &   00:06:04.73 & -29:57:06.7  &  15.74$\pm$0.051 &   0.92$\pm$0.076 &   0.92$\pm$0.117 & 29.33 & -2.74 & ...   &  ...   &  0.4 & -12.8 & 00 &  N  \\
 ZS182                   &   00:06:16.35 & -30:05:57.1  &  11.66$\pm$0.002 &   0.68$\pm$0.004 &   0.61$\pm$0.004 & 29.27 & -4.34 & ...   &  ...   & 21.7 &   3.0 & 96 &  Y  \\
 ZS184                   &   00:06:23.70 & -29:52:04.7  &  15.58$\pm$0.049 &   2.91$\pm$0.050 &         ...      & 29.75 & -3.31 & ...   &  ...   & 38.2 & -78.1 & 00 &  N  \\ 
\enddata
\tablenotetext{a}{Naming convention is as follows: ZS from \citet{deEpstein1985}, BLX from M99 and, PMS04 from P04.}
\tablenotetext{b}{Coordinates are taken from the optical counterpart given in J09 and are J2000.}
\tablenotetext{c}{Intrinsic vales are derived using $E(B-V)=0.016$ and $E(V-I_{c})=0.02$.}
\tablenotetext{d}{Distance used for luminosity calculation is 240 parsecs (see \S \ref{xraylxlbol}); unabsorbed luminosities are in erg s$^{-1}$.}
\tablenotetext{e}{Proper motions and probablities are taken from \citet{Platais2009}; proper motions are in mas yr$^{-1}$.}
\tablenotetext{f}{Membership based on new proper motions. Y$=$proper motion member, N$=$proper motion non-member, NA$=$not available due to limiting magnitude of proper motion survey.}
\tablenotetext{g}{Identical optical counterparts that are identified as two seperate X-ray sources in M99 and P04 (see Table \ref{tab2}).}
\tablenotetext{\dagger}{Listed as single-line spectroscopic binary in \citet{Mermilliod2008}.}
\tablenotetext{\dagger \dagger}{Listed as double-line spectroscopic binary in \citet{Mermilliod2008}.}
\tablenotetext{\ddagger}{Listed as a single-line spectroscopic binary in \citet{Jeffries1999a}.}
\end{deluxetable}

 \clearpage

 \begin{deluxetable}{l c c l c c c c}
\tablecolumns{8}
\tablewidth{0pt}
\tabletypesize{\scriptsize}
\tablecaption{Multiple X-ray Source Detection to Single Optical Counterparts\label{tab2}}
\tablehead{
\multicolumn{3}{c}{{\sc rosat}} & \multicolumn{3}{c}{{\sc xmm-newton}} &  \multicolumn{2}{c}{Optical Counterpart} \\
\colhead{ID} & \colhead{RA} & \colhead{DEC} & \colhead{ID} & \colhead{RA} & \colhead{DEC} & \colhead{RA} & \colhead{DEC}
}
\startdata
ZS44   &  0:02:14.5 &  -29:48:58.6 &  PMS04-182 &  00:02:14.6 &  -29:49:03.5 &  00:02:14.65 & -29:49:04.41 \\
ZS60   &  0:02:41.8 &  -29:58:53.7 &  PMS04-97  &  00:02:41.7 &  -29:58:55.8 &  00:02:41.73 & -29:58:53.19 \\
BLX-12 &  0:02:07.2 &  -30:04:41.8 &  PMS04-48  &  00:02:07.6 &  -30:04:44.4 &  00:02:07.65 & -30:04:42.90 \\
BLX-15 &  0:02:22.3 &  -30:02:52.0 &  PMS04-66  &  00:02:22.7 &  -30:02:52.3 &  00:02:22.87 & -30:02:52.88 \\
BLX-16 &  0:02:23.1 &  -29:50:34.5 &  PMS04-169 &  00:02:23.4 &  -29:50:40.4 &  00:02:23.58 & -29:50:39.99 \\
BLX-42 &  0:03:22.3 &  -29:53:49.4 &  PMS04-150 &  00:03:22.6 &  -29:53:52.2 &  00:03:22.73 & -29:53:50.89 \\
\enddata
\tablecomments{Coordinates are J2000.0 Equinox.}
\end{deluxetable}

 \clearpage

 \begin{deluxetable}{c c c}
\tablecolumns{3}
\tablewidth{0pt}
\tabletypesize{\scriptsize}
\tablecaption{Mean values of $L_{x}$ for Blanco~1\label{tab3}}
\tablehead{
\colhead{Spectral Range\tablenotemark{a}}    & \colhead{Mean $Log(L_{x})$}  & \colhead{$1\sigma$}\\
                                    & [erg s$^{-1}$]               & [erg s$^{-1}$]\\}
\startdata
ROSAT - $B-V$ \\
\hline
G and Earlier & 29.52 & 0.21 \\
K             & 29.66 & 0.30 \\
M and Later   & 29.61 & 0.52 \\
\hline
ROSAT - $V-I_{c}$\\
\hline
G and Earlier & 29.57 & 0.22 \\
K             & 29.65 & 0.33 \\
M and Later   & 29.23 & 0.21 \\
\hline
XMM:Newton - $B-V$\\
\hline
G and Earlier & 29.60 & 0.31 \\
K             & 29.86 & 0.37 \\
M and Later   & 29.69 & 0.40 \\
\hline
XMM:Newton - $V-I_{c}$\\
\hline
G and Earlier & 29.62 & 0.33 \\
K             & 29.76 & 0.47 \\
M and Later   & 29.41 & 0.33 \\
\enddata
\tablenotetext{a}{Spectral ranges are defined by \citet{Kenyon1995}: G-type and Earlier $B-V_{0} < 0.8$, K-type $0.8 < B-V_{0} < 1.3$, M-type and later $B-V_{0} > 1.3$}
\tablecomments{X-ray luminosities were measured over the energy ranges of 0.1-2.4 keV and 0.3-5.0 keV for {\sc rosat} and {\sc xmm}-Newton, respectively.}
\end{deluxetable}

 \clearpage

 \begin{deluxetable}{l c c c c c}
\tablecolumns{6}
\tablewidth{0pt}
\tabletypesize{\scriptsize}
\tablecaption{$L_{x}$ and $L_{x}/L_{bol}$ for Several Open Clusters\label{tab4}}
\tablehead{
\colhead{Name}  & \colhead{Reference}    & \colhead{$\#$ of Stars} & \colhead{$Log($Age$)$} & \colhead{Mean $Log(L_{x})$}  & \colhead{Mean} \\ 
\colhead{}      & \colhead{Number\tablenotemark{b}} & \colhead{Included}   & \colhead{[yr]}        & \colhead{[erg s$^{-1}$]}     & \colhead{$Log(L_{x}/L_{bol})$}
}
\startdata
G Spectral Type\\
and Earlier\tablenotemark{a} \\
\hline
NGC~2547     & 1 & 22  & 7.48 &  29.71$\pm$0.46 & -4.30$\pm$0.73\\
IC~2602\tablenotemark{c} & 2 & 9  & 7.56 &  ...           & ... \\
IC~2391      & 3 & 20  & 7.56 &  29.76$\pm$0.51 & -4.64$\pm$1.14\\
$\alpha$~Persei & 4 & 30 & 7.72 &  29.72$\pm$0.49 & -3.67$\pm$0.57\\
Blanco~1\tablenotemark{d}   & 5 & 11 & 7.90 &  29.60$\pm$0.28 & -4.29$\pm$0.24 \\
Pleiades     & 6 & 31 & 8.08 &  29.44$\pm$0.36 & -3.94$\pm$0.41 \\
NGC~2516     & 7 & 54 & 8.15 &  29.40$\pm$0.42 & -4.63$\pm$0.56 \\
M~7          & 8 & 47 & 8.35 &  29.52$\pm$0.30 & -4.32$\pm$0.32 \\
Hyades       & 9 & 94 & 8.80 &  28.98$\pm$0.32 & -4.96$\pm$0.60 \\
Praesepe     & 10 & 36 & 8.86 &  29.18$\pm$0.38 & -4.82$\pm$0.53 \\
\hline
K Spectral Type\tablenotemark{a} \\
\hline
NGC~2547     & 1 & 23 & 7.48 &  29.92$\pm$0.37 & -3.38$\pm$0.30 \\
IC~2602      & 2 & 16 & 7.56 &  29.67$\pm$0.43 & -3.54$\pm$0.30 \\
IC~2391      & 3 & 13 & 7.56 &  29.87$\pm$0.22 & -3.32$\pm$0.15 \\
$\alpha$~Persei & 4 & 51 & 7.72 &  29.67$\pm$0.42 & -3.27$\pm$0.38 \\
Blanco~1\tablenotemark{d} & 5 & 25 & 7.90 &  29.71$\pm$0.40 & -3.32$\pm$0.24 \\
Pleiades     & 6 & 40 & 8.08 &  29.44$\pm$0.29 & -3.21$\pm$0.29 \\
NGC~2516     & 7 & 81 & 8.15 &  29.30$\pm$0.47 & -3.77$\pm$0.55 \\
M~7          & 8 & 56 & 8.35 &  29.42$\pm$0.32 & -3.55$\pm$0.44 \\
Hyades       & 9 & 42 & 8.80 &  28.83$\pm$0.53 & -4.52$\pm$0.81 \\
Praesepe     & 10 & 12 & 8.86 &  29.40$\pm$0.43 & -4.10$\pm$0.78 \\
\hline
M Spectral Type\\
and Later\tablenotemark{a}\\
\hline
NGC~2547     & 1 & 52 & 7.48 &  29.42$\pm$0.32 & -3.06$\pm$0.31 \\
IC~2602      & 2 & 27 & 7.56 &  29.15$\pm$0.27 & -3.27$\pm$0.29 \\ 
IC~2391\tablenotemark{c} & 3 &  4 & 7.56 &  ...     & ...       \\
$\alpha$~Persei & 4 & 50 & 7.72 &  29.24$\pm$0.33 & -2.89$\pm$0.34 \\
Blanco~1\tablenotemark{d} & 5 & 11 & 7.90 &  29.33$\pm$0.27 & -3.16$\pm$0.24 \\
Pleiades     & 6 & 43 & 8.08 &  29.16$\pm$0.20 & -2.90$\pm$0.23 \\
NGC~2516     & 7 & 90 & 8.15 &  28.97$\pm$0.31 & -2.98$\pm$0.34 \\
M~7\tablenotemark{b} & 8 &  4 & 8.35 &  ...    & ...            \\
Hyades       & 9 & 48 & 8.80 &  28.84$\pm$0.36 & -3.31$\pm$0.40 \\
Praesepe     & 10 & 20 & 8.86 &  29.09$\pm$0.28 & -3.11$\pm$0.47 \\
\enddata
\tablenotetext{a}{Spectral ranges are defined by \citet{Kenyon1995}: G-type and Earlier $B-V_{0} < 0.8$, K-type $0.8 < B-V_{0} < 1.3$, M-type and later $B-V_{0} > 1.3$}
\tablenotetext{b}{The letters that come after the following citations represent -X {\sc xmm}-Newton[0.3-10 keV] and -R {\sc rosat}[0.1-2.4 leV]. {\it References:}{\bf (1)}\citealt{Jeffries2006}-X, {\bf (2)}\citealt{Randich1995a}-R, {\bf (3)}\citealt{Patten1996}-R,{\bf (4)}\citealt{Randich1996}-R, {\bf (5)}Our new analysis, {\bf (6)}\citealt{Stauffer1994,Micela1999b}-R, {\bf (7)}\citealt{Pillitteri2006}-X, {\bf (8)}\citealt{Prosser1995b}-R, {\bf (10)}\citealt{Stern1995}-R.,\citealt{Perryman1998}, {\bf (9)}\citealt{Randich1995b}-R ({\bf {\em N.B.}:} These reference numbers are also used in Fig.\ \ref{fig8}.)}
\tablenotetext{c}{Did not include mean $L_{x}$ or $L_{x}/L_{bol}$ values due to insufficent X-ray detections in mass ranges.}
\tablenotetext{d}{For stars observed with {\sc rosat} and {\sc xmm-newton}, we use the mean values in the calculations.}
\end{deluxetable}

 \end{document}